\documentclass{jfm}
\usepackage{graphicx}
\usepackage{epstopdf, epsfig}

\usepackage{float}

\usepackage{amsmath}
\usepackage[colorlinks, citecolor=blue]{hyperref}
\usepackage{siunitx}
\usepackage[usenames,dvipsnames]{xcolor}
\usepackage{setspace}

\usepackage{lineno}
\usepackage{lipsum}

\newcommand{\DL}[1]{{\textcolor{black}{#1}}}
\newcommand{\VS}[1]{{\textcolor{black}{#1}}}
\newcommand{\AO}[1]{{\textcolor{black}{#1}}}
\newcommand{\KZ}[1]{{\textcolor{black}{#1}}}
\newcommand{\AKD}[1]{{\textcolor{black}{#1}}}
\newcommand{\rev}[1]{{\textcolor{black}{#1}}}

\usepackage[textwidth=\dimexpr\textwidth-2cm\relax]{todonotes}
\makeatletter
\@mparswitchfalse%
\makeatother
\normalmarginpar %for right-handed notes and lines, or

\newcommand{\oo}{\color{black} \normalfont}
\newcommand{\ooT}{\color{black} \normalfont}
\newcommand{\bb}{\color{black} \normalfont}

\shorttitle{Viscoelastic Worthington jets \& droplets produced by bursting bubbles}

\shortauthor{A. K. Dixit, A. Oratis, K. Zinelis, D. Lohse \& V. Sanjay}
\title{Viscoelastic Worthington jets \& droplets produced by bursting bubbles}

\author{Ayush K. Dixit\aff{1} \corresp{\email{a.k.dixit@utwente.nl}},
	Alexandros Oratis\aff{1}  \corresp{\email{a.oratis@utwente.nl}},
	Konstantinos Zinelis\aff{2,3}, \corresp{\email{k.zinelis17@imperial.ac.uk}}
	Detlef Lohse\aff{1,4} \corresp{\email{d.lohse@utwente.nl}},
	\and Vatsal Sanjay\aff{1} \corresp{\email{vatsalsanjay@gmail.com}}}

\affiliation{\aff{1} Physics of Fluids Group, Max Planck Center for Complex Fluid Dynamics, Department of Science and Technology, and J. M. Burgers Centre for Fluid Dynamics, University of Twente, P. O. Box 217, 7500 AE Enschede, The Netherlands
\aff{2} Department of Chemical Engineering, Imperial College London, London SW7 2AZ, UK
\aff{3} Department of Chemical Engineering, Massachusetts Institute of Technology, Cambridge, MA 02139, USA
\aff{4} Max Planck Institute for Dynamics and Self-Organization,  Am Fassberg 17, 37077 G\"ottingen, Germany}

\begin{document}
\maketitle

\begin{abstract}
Bubble bursting and subsequent collapse of the open cavity at free surfaces of contaminated liquids can generate aerosol droplets, facilitating pathogen transport. After film rupture, capillary waves focus at the cavity base, potentially generating fast Worthington jets that are responsible for ejecting the droplets away from the source. While extensively studied for Newtonian fluids, the influence of non-Newtonian rheology on this process remains poorly understood.
\oo
Here, we employ direct numerical simulations to investigate the bubble cavity collapse in viscoelastic media, such as polymeric liquids.
We find that the jet and drop formation are dictated by two dimensionless parameters: the elastocapillary number $Ec$ (the ratio of the elastic modulus and the Laplace pressure) and the Deborah number $De$ (the ratio of the relaxation time and the inertio-capillary timescale).
We show that for low values of $Ec$ and $De$, the viscoelastic liquid adopts a Newtonian-like behavior, where the dynamics are governed by the solvent Ohnesorge number $Oh_s$ (the ratio of visco-capillary and inertio-capillary timescales).
In contrast, for large values $Ec$ and $De$, the enhanced elastic stresses completely suppress the formation of the jet.
\bb
For some cases with intermediate values of \oo$Ec$ and $De$\bb, smaller droplets are produced compared to Newtonian fluids, potentially enhancing aerosol dispersal. By mapping the phase space spanned by \oo$Ec$, $De$, and $Oh_s$\bb, we reveal three distinct flow regimes: (i) jets forming droplets, (ii) jets without droplet formation, and (iii) absence of jet formation. Our results elucidate the mechanisms underlying aerosol suppression versus fine spray formation in polymeric liquids, with implications for pathogen transmission and industrial processes involving viscoelastic fluids.
\end{abstract}

%\begin{keywords}
%Authors should not enter keywords on the manuscript, as these must be chosen by the author during the online submission process and will then be added during the typesetting process (see http://journals.cambridge.org/data/\linebreak[3]relatedlink/jfm-\linebreak[3]keywords.pdf for the full list)
%\end{keywords}

\section{Introduction}
\label{sec:Intro}
\begin{figure}
	\centering
	\includegraphics[width=\textwidth]{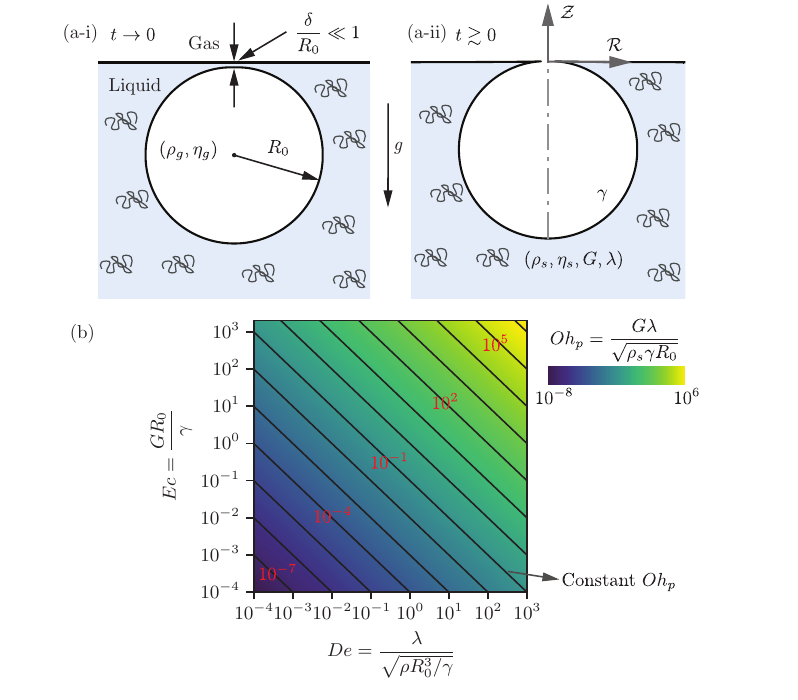}
	\caption { \AKD{(a-i)} A bubble with radius $R_0$ rests \DL{close to} the liquid-gas interface, separated \DL{from it} by a thin liquid film of thickness $\delta \ll R_0$. The surrounding viscoelastic medium is characterized by density $\rho_s$, solvent viscosity $\eta_s$, elastic modulus $G$, and relaxation time $\lambda$. The gas has density $\rho_g$ and viscosity $\eta_g$. \AKD{(a-ii)} Film rupture creates an axisymmetric cavity, which we study in this work. (b) Apart from the solvent Ohnesorge number $Oh_s = \eta_s/\sqrt{\rho_s\gamma R_0}$ and the Bond number $Bo = \rho_sgR_0^2/\gamma$, the presence of polymers introduces two additional parameters, \DL{namely} the elastocapillary number $Ec = GR_0/\gamma$ (equation~\eqref{eq:Ecdef}) and the Deborah number $De = \lambda/\sqrt{\rho_s R_0^3/\gamma}$ (equation~\eqref{eq:Dedef}). To \DL{explore} the dynamics, we move across the entire $Ec$-$De$ phase space. Often, the polymeric Ohnesorge number $Oh_p = G\lambda/\sqrt{\rho_s\gamma R_0} \DL{ = Ec \times De}$ (equation~\eqref{eq:Ohpdef})  based on polymeric viscosity is also used to describe the influence of polymers.}
	\label{schematic}
\end{figure}

%\vsy{\#TODO-DONE: add the three experimental papers, \citet{cheny1996extravagant, rodriguez2023bubble, cabalganteeffect} if not already cited!!! and find the relevant parameter of interest from experiments.}
Bubbles in liquids \citep{Lohse2018} -- from oceans \citep{deike2022mass} and volcanoes \citep{gonnermann2007fluid} to cosmetic gels \citep{lin1970mechanisms, daneshi2024growth} and champagne \citep{liger2012physics,mathijssen2023culinary} -- rise due to buoyancy and reach the liquid-gas interface, where they sit as the intervening liquid film drains \citep[figure~\ref{schematic}a-i,][]{lhuissier2012bursting, bartlett2023universal}.
Upon film rupture, numerous tiny droplets\AO{, known as film droplets,} scatter over the free surface \citep{lhuissier2012bursting,villermaux2022bubbles}, leaving a high-energy bubble cavity \citep[figure~\ref{schematic}a-ii,][]{woodcock1953giant,knelman1954mechanism,mason1954bursting}. The subsequent collapse of this cavity is driven by surface tension. This process involves rim retraction \citep{taylor-1959-procrsoclonda,culick-1960-japplphys,sanjay-2022-JFM} that generates capillary waves \citep{snoeijer2025coalescence}.
These waves propagate along the cavity, converging at its base to create an inertial flow focusing \citep{gordillo2019capillary,gordillo2023theory} that forms a Worthington jet \citep{worthington1877xxviii,worthington1908study,stuhlman1932mechanics,lohse2004impact,VatsalThesis} that features large strain rates \citep{sen2024elastocapillary}. The jet may fragment into droplets through end-pinching and the Rayleigh--Plateau instability \citep{rayleigh1878instability, plateau1873statique, keller1995blob, stone1989relaxation, ghabache2016size, walls2015jet}.
These jet droplets, typically larger and faster than the initial film droplets, play a crucial role in transporting dissolved substances to the atmosphere \citep{berny2020role,villermaux2022bubbles,dubitsky2023enrichment}.
The dynamics of bubble bursting have far-reaching implications across various domains. These include the transfer of pathogens from contaminated water to air \citep{bourouiba2021fluid}, the transport of dissolved salt from seawater to the atmosphere, where salt particles act as cloud condensation nuclei \citep{dubitsky2023effects, de2011production}, and the dynamics in bioreactors containing animal cells \citep{boulton1993gas}. The unique capacity of ejected droplets to transport diverse species underscores the importance of comprehending the complete dynamics \KZ{that dictate} their formation.
Ever since the first documented study of \citet{stuhlman1932mechanics}, advanced experiments and simulations have extensively characterized the rich dynamics of bursting bubbles.
Key metrics include ejected drop heights \citep{stuhlman1932mechanics}, sizes \citep{kientzler1954photographic,deike2018dynamics,berny2020role,berny2021statistics,blanco2020sea,villermaux2022bubbles}, and velocities \citep{deike2018dynamics, gordillo2019capillary, sanjay2021bursting, gordillo2023theory}.

\citet{macintyre1972flow} revealed internal liquid flow using dye and attempted to understand the drop composition, which was finally explained by direct numerical simulations (DNS) of \citet{dubitsky2023enrichment}.
Furthermore, \citet{dasouqi2022effect} demonstrated atmospheric flow patterns using smoke-filled bubbles, which were detailed numerically by \citet{singh2021dynamics}.
Although shadowgraphy techniques limit most experimental studies, x-ray imaging has captured traveling capillary wave dynamics, providing crucial validation for DNS results \citep{lee2011size}. These advancements have significantly enhanced our understanding of bubble bursting at the Newtonian liquid-gas interface across various scales and applications.
Indeed, for a bubble of radius $R_0$ surrounded by a liquid with viscosity, density, and surface tension $\eta_s$, $\rho_s$, and $\gamma$, the interplay of capillarity, viscosity, and gravity governs the bubble cavity collapse. \DL{Correspondingly, the key control parameters of this process are} the solvent Ohnesorge number

\begin{align}
	\label{eq:Ohdef}
	Oh_s = \frac{\eta_s}{\sqrt{\rho_s\gamma R_0}},
\end{align}

\noindent and the Bond number

\begin{align}
	\label{eq:Bodef}
	Bo = \frac{\rho_sgR_0^2}{\gamma}.
\end{align}

\noindent Here, $g$ is the acceleration due to gravity. The solvent Ohnesorge number $Oh_s$ exemplifies the dimensionless viscosity of the surrounding medium, significantly influencing the capillary wave dynamics, determining their damping and overall viscous dissipation, while \DL{the Bond number} $Bo$ affects the initial cavity shape and \DL{the} hydrostatic pressure differences \citep{walls2015jet,bergmann2006giant,bergmann2009controlled,Lohse2018}. In this study, we will focus our attention on the limiting case of very small \DL{bubbles} with $Bo = 0.001$, for which the bubbles can be approximated as spheres \citep[figures~\ref{schematic}a,][]{toba1959drop,princen1963shape,lhuissier2012bursting}.
For the Newtonian cases, appendix~\ref{app:newtonian_limit} summarizes the key results, including the effect of $Oh_s$ on bubble-busting dynamics. For the influence of gravity on the shape and consequently the overall dynamics of \oo Newtonian fluids,\bb\, we refer the readers to \citet{toba1959drop,princen1963shape,walls2015jet,krishnan2017scaling,deike2018dynamics}.

Given the potential for jet drops to transport pathogens or pollutants into the atmosphere, strategies to prevent their generation are pertinent. Recent studies unsurprisingly show that non-Newtonian effects, particularly that \oo viscoplasticity and\bb\, viscoelasticity, can suppress jet drop production \citep{sanjay2021bursting, sen2021retraction, rodriguez2023bubble, ji2023secondary}.
While computational studies have successfully reproduced experimental observations, such as elasticity-induced droplet suppression \citep{cabalganteeffect, ari2024bursting}, the full impact of these effects on bubble-bursting dynamics remains elusive. \oo In this paper, we answer the question: How does the viscoelasticity influence the observed regimes? What underlying physics governs the transitions between these regimes?\bb\,
Advancements in solving non-linear constitutive equations for highly deformed interfacial flows of \oo viscoelastic fluids\bb\, have been made possible by techniques like the log-conformation method \citep{fattal2004constitutive} and the square-root conformation method \citep{balci2011symmetric}.
Originally developed for single-phase flows, these methods have been extended to multiphase flows \citep{fraggedakis2016velocity,lopez2019adaptive,varchanis2022numerical,francca2024elasto,zinelis2023transition},
facilitating more comprehensive investigations into this topic.

Viscoelastic media differ from viscous Newtonian liquids in their rheological properties, exhibiting both viscous and elastic stresses when deformed due to the presence of dissolved polymers.
These polymeric effects are characterized by two material properties: \DL{the} elastic modulus $G$ that characterizes the strength of \DL{the dissolved} polymers by relating the strain with the additional polymeric stresses in the system, and \DL{the} relaxation time scale $\lambda$ that characterizes the memory of the system as it is a measure of the timescale at which the additional polymeric stresses in the system \AO{vanish}. \DL{When non-dimensionalizing these properties, we obtain two further non-dimensionalized control parameters, namely,} the elastocapillary number

\begin{align}
	\label{eq:Ecdef}
	Ec = \frac{GR_0}{\gamma},
\end{align}

\noindent comparing the elastic modulus to the Laplace pressure scale\DL{,} and the Deborah number

\begin{align}
	\label{eq:Dedef}
	De = \frac{\lambda}{\sqrt{\rho_sR_0^3/\gamma}},
\end{align}

\noindent comparing the relaxation time of the additional stresses to the process timescale, i.e., the inertiocapillary timescale $\tau_{\gamma} = \sqrt{\rho_sR_0^3/\gamma}$. Additionally, we also introduce the polymeric viscosity $\eta_p = G\lambda$ based on dimensional arguments, which can be normalized with the inertiocapillary scales to give the polymeric Ohnesorge number (figure~\ref{schematic}b)

\begin{align}
	\label{eq:Ohpdef}
	Oh_p = \frac{\eta_p}{\sqrt{\rho_s\gamma R_0}} = \DL{Ec \times De},
\end{align}

\noindent which is the product of $Ec$ and $De$. \oo We note here that $Oh_p$ and $Oh_s$ are related by

\begin{align}
	Oh_p = \dfrac{\eta_p}{\eta_s} \, Oh_s = c\,Oh_s,
\end{align}

\noindent where $c = \eta_p/\eta_s$ is the so-called concentration of the polymers (see e.g., \citet{remmelgas1999computational, hinch2024fast}). \bb

\oo Prior experimental studies have provided valuable insights into viscoelastic effects on bubble bursting dynamics. Early work by \citet{cheny1996extravagant} demonstrated dramatic modifications of Worthington jets through polymer addition, where even small concentrations ($c \sim 50$ ppm) reduced jet heights by an order of magnitude.
More recently, \citet{rodriguez2023bubble} demonstrated how even weakly viscoelastic polymer solutions (with relaxation times $\lambda \leq 50\si{\micro\second}$) can dramatically alter bubble bursting dynamics through both interfacial and bulk effects. They found that at optimal polymer concentrations ($\approx 25\,\text{ppm}$), interfacial effects enhanced jet velocity by dampening short-wavelength capillary waves, while at higher concentrations, extensional thickening led to complete droplet suppression.
The elastic stress buildup during jet formation was further elucidated by \citet{cabalganteeffect}, who supported the previous observation that droplet emission is completely suppressed for large enough relaxation times (jet Weissenberg number $Wi_j = \lambda v_j/R \ge 0.5$, where $v_j$ is the characteristic velocity of the Worthington jet), while the jet velocity is primarily dictated by $Oh_p$. These experimental observations motivate our systematic computational investigation of the $Oh_s$-$Ec$-$De$ phase space to uncover the fundamental mechanisms that govern viscoelastic bubble bursting. We refer readers to appendix~\ref{app:accounting} for a representative summary of the different control parameters.\bb

In this study, we investigate viscoelastic effects on bubble bursting dynamics by \DL{exploring} the three-dimensional phase space of $Oh_s$, $Ec$, and $De$, using volume of fluid-based finite volume simulations.
\oo Using the Oldroyd-B constitutive relation, we demonstrate\bb\, that the addition of polymers significantly influences the overall dynamics, which are governed by the interplay of viscous and elastic effects.
\VS{For systems with a permanent memory of its initial state and subsequent deformations, i.e., when the additional polymeric stresses are sustained throughout the process time scale ($De \to \infty$), the dimensionless elastic modulus dictates the dynamics and suppression of jet and drops.
In contrast, for systems with poor memory of its initial state and subsequent deformation ($De \to 0$), the dynamics resemble those encountered in Newtonian liquids with an effective viscosity deduced using the slender elastic jet equations.}
\oo Despite its simplicity, we note that Oldroyd-B model has some crucial limitations.
For instance, it cannot account for the shear-thinning behavior of polymer solutions and it predicts the divergence of stresses for strong extensional flows \citep{yamani2023master,alves2021numerical}.  
Consequently, the Oldroyd-B model cannot accurately capture the final stages of filament thinning or the actual rupture of viscoelastic filaments, which may affect predictions of droplet detachment and fine aerosol formation.
Nevertheless, we choose the Oldroyd-B model as its simplicity allows us to gain fundamental insight into the interplay between capillary, viscous, and elastic forces during bubble bursting.  \bb\,

Building upon the extensive literature on viscoelastic flows, we extend these concepts to the specific case of bubble bursting.
Previous research has explored viscoelastic phenomena in various contexts, including flow through nozzles and contractions \citep{hinch1993flow, chen1991interfacial, boyko2024flow}, stability and breakup of viscoelastic jets \citep{middleman1965stability, goren1982surface, bousfield1986nonlinear, chang1999iterated, anna2001elasto,pandey2021elastic,sen2024elastocapillary, zinelis2023transition}, coalescence and spreading of viscoelastic drops and bubbles \citep{bouillant2022rapid, dekker2022elasticity, oratis2023coalescence}, and oscillating bubbles in viscoelastic media \citep{oratis2024unifying}.
Recent studies have also investigated elastoviscoplastic flows, incorporating viscous, elastic, and plastic aspects \citep{putz2009solid, varchanis2019modeling, francca2024elasto, ari2024bursting}, further expanding our understanding of non-Newtonian liquids.
We refer readers to reviews by \citet{bogy1979drop}, \citet{eggers1997nonlinear}, and \citet{yarin1993free} for comprehensive overviews of these topics.
Our work applies the foundational knowledge developed in these works to elucidate how viscoelasticity alters the formation of Worthington jets and ejected droplets during bubble bursting, enhancing our understanding of this specific phenomenon.

This paper is organized as follows: \S~\ref{sec:method} presents the governing equations and numerical method. \VS{\S~\ref{sec:polymers} investigates the polymer influence on bubble bursting, focusing on systems with permanent ($De \to \infty$, \S~\ref{sec:LargeDe}) and poor ($De \to 0$, \S~\ref{sec:shortmemory}) memory. For both cases, we categorize bursting bubble dynamics into distinct regimes and elucidate the transitions in \S~\ref{sec:regimes} where we generalize the results across systems where the memory of the initial conditions and subsequent deformations is gradually fading ($0 < De < \infty$).}
Finally, \S~\ref{sec:Conclusion} summarizes our findings and suggests future research directions.

\section{Numerical framework and problem description}
\label{sec:method}

\subsection {Governing Equation}
\label{subsec:governingEqn}

We investigate the collapse of an open bubble cavity at the interface in a viscoelastic medium \DL{(of figure \ref{schematic})} using an axisymmetric domain with incompressible fluids.
Length scales are normalized using the initial bubble radius giving $\mathcal{L} = \tilde{\mathcal{L}}R_0$ \DL{as characteristic length,} and the time is normalized using \DL{the} \AKD{inertiocapillary} timescale $\tau_\gamma = \sqrt{\rho_s {R_0}^3/\gamma}$ giving $t = \tilde{t}\tau_{\gamma}$. These normalizations yield an \AKD{inertiocapillary} velocity scale $u_{\gamma} = \sqrt{\gamma/ \rho_{s} R_0}$ for the velocity field $\boldsymbol{u} = \tilde{\boldsymbol{u}}u_\gamma$. Lastly, all stresses are normalized using the Laplace pressure scale, $\boldsymbol{\sigma} = \tilde{\boldsymbol{\sigma}}\sigma_\gamma$, where $\sigma_\gamma = \gamma/R_0$.
\DL{Here, as usual, non-dimensionalized quantities are denoted with a tilde, though from here onwards, we drop the tilde, and all equations are thus dimensionless in the current section.}
Throughout the manuscript, we use the subscripts $s$, $p$, and $g$ to denote liquid solvent, polymer, and gas, respectively.
The governing mass and momentum conservation equations for the liquid phase read as

\begin{align}
	\label{massconserve}
	\boldsymbol{\nabla\cdot u}=0,\,\text{and}
\end{align}

\begin{align}
	\label{momconserve}
	\frac{\p \boldsymbol{u}}{\p t} + \boldsymbol{\nabla\cdot} \left(\boldsymbol{u}\boldsymbol{u}\right) =  -\boldsymbol{\nabla}p + \boldsymbol{\nabla\cdot}\left(\boldsymbol{\sigma_s}+\boldsymbol{\sigma_p}\right),
\end{align}

\noindent where the Newtonian contribution (coming from the solvent) $\boldsymbol{\sigma_s}$ is

\begin{align}
	\label{taus}
	\boldsymbol{\sigma_{s}} =  2 Oh_s \boldsymbol{\mathcal{D}},
\end{align}

\noindent with $\boldsymbol{\mathcal{D}} = \left(\boldsymbol{\nabla u} + \left( \boldsymbol{ \nabla u} \right) ^T \right)/2$ representing the symmetric part of the velocity gradient tensor\oo--equal to half of the rate-of-strain tensor.\bb\,
The non-Newtonian contribution $\boldsymbol{\sigma_{p}}$ arises from the presence of polymers in the fluid. We emphasize that although we refer to $\boldsymbol{\sigma_{p}}$ as `polymeric stresses' in the context of dilute polymer liquids, this concept extends to any deformable microstructure within the fluid that responds to flow \citep{saramito2007,snoeijer2020relationship,francca2024elasto,ari2024bursting}.
To characterize the deformation of these microstructures, we introduce the conformation tensor $\boldsymbol{\mathcal{A}}$, an order parameter that evolves from an initial identity state $\boldsymbol{\mathcal{A}} = \boldsymbol{\mathcal{I}}$ (figure~\ref{schematic}a-ii).
\oo Here, we employ the Oldroyd-B model, which represents the simplest conformation tensor-based constitutive equation for viscoelastic fluids \bb\,\citep{oldroyd1950formulation, bird1977dynamics, snoeijer2020relationship, stone2023note, boyko2024perspective}. This model assumes a linear relationship between elastic stresses and polymeric deformation,

\begin{align}
	\label{sigmap}
	\boldsymbol{\sigma_{p}} = Ec \left(\boldsymbol{\mathcal{A}} - \boldsymbol{\mathcal{I}}\right),
\end{align}

\noindent where $Ec$ is the elastocapillary number (equation~\eqref{eq:Ecdef}), representing the strength of the polymers analogous to a dimensionless elastic modulus. \AO{Note that even though the polymeric stresses $\boldsymbol{\sigma_{p}}$ grow linearly with $\boldsymbol{\mathcal{A}}$, the polymeric deformations $\boldsymbol{\mathcal{A}}$ can be highly nonlinear}. Naturally, in the limit of $Ec = 0$, the polymeric stress would vanish, and the system will give a viscous Newtonian dictated by the solvent Ohnesorge number $Oh_s$ (see equation~\eqref{taus}).

Additionally, the conformation tensor $ \boldsymbol{\mathcal{A}}$ relaxes to its base state $\boldsymbol{\mathcal{I}}$ over time due to thermal effects.
Once more, using the Oldroyd-B model, $ \boldsymbol{\mathcal{A}}$ follows a linear relaxation law 
\oo (i.e., the rate of change of $\boldsymbol{\mathcal{A}}$ in the Lagrangian frame is linear in $\boldsymbol{\mathcal{A}}$),\bb\,

\begin{align}
	\label{Aupperconv}
	\stackrel{\smash{\raisebox{0ex}{$\mkern8mu\boldsymbol{\nabla}$}}}{\boldsymbol{\mathcal{A}}}  =  - \frac{1}{De} \left( \boldsymbol{\mathcal{A}} - \boldsymbol{\mathcal{I}}  \right),
\end{align}

\noindent where

\begin{align}
    \label{Aupper_def}
    \stackrel{\smash{\raisebox{0ex}{$\mkern8mu\boldsymbol{\nabla}$}}}{\boldsymbol{\mathcal{A}}} \equiv \frac{\partial\boldsymbol{\mathcal{A}}}{\partial t} + \left(\boldsymbol{u\cdot\nabla}\right)\boldsymbol{\mathcal{A}} - \boldsymbol{\mathcal{A}\cdot}\left(\boldsymbol{\nabla u}\right) - \left(\boldsymbol{\nabla u}\right)^T\boldsymbol{\cdot\mathcal{A}}
\end{align}

\noindent is the frame-invariant upper convected Oldroyd derivative of second-rank tensor $\boldsymbol{\mathcal{A}}$, and $De = \lambda/\tau_\gamma$ (defined in equation~\eqref{eq:Dedef}) is the Deborah number, representing the ratio of the polymer relaxation time $\lambda$ to the process timescale $\tau_\gamma$. \oo We note that while the Oldroyd-B model is nonlinear in terms of the velocity field and its gradient, both the stress term and its relaxation law remain linear in $\boldsymbol{\mathcal{A}}$. This characteristic contrasts with models such as the Giesekus model, which involves a quadratic term $\boldsymbol{\mathcal{A}\cdot\mathcal{A}}$ \citep{giesekus1982simple}, or the FENE models, which include a nonlinear term involving a finite-extensibility parameter $L$ \citep{bird1980polymer}. Therefore, the Oldroyd-B model is often referred to as ``quasi-linear”  \citep{davoodi2018secondary, alves2021numerical}\bb.

The Deborah number characterizes \DL{the polymeric liquid's} memory.
It is instructive to note that in the limit of $De \to \infty$, \DL{polymeric liquids} have permanent memory and the dissolved polymers undergo affine motion \citep[see equation~\eqref{Aupperconv} and][]{snoeijer2020relationship,stone2023note,boyko2024perspective}

\begin{align}
	\label{eq:affineMotion}
	\stackrel{\smash{\raisebox{0ex}{$\mkern8mu\boldsymbol{\nabla}$}}}{\boldsymbol{\mathcal{A}}} = 0,
\end{align}

\noindent indicating that they follow the flow and deform according to the velocity field. In this limit, for finite $Ec$, the Oldroyd-B model is equivalent to the damped neo-Hookean model (also known as the Kelvin-Voigt model) for solids \citep{snoeijer2020relationship}.
Conversely, at $De = 0$, \DL{polymeric liquids} have no memory of their initial condition and subsequent deformations, relaxing immediately to the base state. For non-infinite $Ec$ values, polymeric stresses vanish, resulting in a Newtonian response (equation~\eqref{sigmap}) governed by the solvent Ohnesorge number $Oh_s$ (see equation~\eqref{taus}).
\AO{It is, therefore, surprising that} both $Ec = 0$ and $De = 0$ (figure~\ref{schematic}b) represent Newtonian responses, \DL{irrespectively of the corresponding other parameter.}

Equations \eqref{sigmap} and \eqref{Aupperconv} \AO{can be} combined to get

\begin{align}
	\label{oldroydb}
	De\stackrel{\smash{\raisebox{0ex}{$\mkern-8mu\boldsymbol{\nabla}$}}}{\boldsymbol{\sigma_p}} +\,\,\boldsymbol{\sigma_p}
	= 2Oh_p {\boldsymbol{\mathcal{D}}},
\end{align}

\noindent where $Oh_p = Ec \times De$ is the polymeric Ohnesorge number (equation~\eqref{eq:Ohpdef}). Consequently, in the limit $De \to 0$ at fixed $Oh_p$ (e.g., moving along constant $Oh_p$ lines in figure~\ref{schematic}b), the system exhibits a viscous Newtonian response with a total dimensionless viscosity of $Oh_s+Oh_p$.

The Oldroyd-B model, despite its widespread use due to \AO{its} simplicity, fails to capture several important physical phenomena \citep{snoeijer2020relationship}. It is inadequate to describe shear-thinning behavior in polymeric liquids \citep{yamani2023master} and erroneously predicts unbounded stress growth in strong extensional flows \citep{mckinley2002filament, eggers2020self}. \ooT The numerical discretization of Oldroyd-B (\S~\ref{sec:methods}) also features an implicit stress regularization due to the finite grid size \citep{renardy2021mathematician}--similar in sprit to the implicit slip regularization of the contact line singularity \citep{afkhamiTransitionNumericalModel2018,fullanaConsistentTreatmentDynamic2024}.\bb
These limitations can be addressed by incorporating finite polymer extension, for example, by increasing the effective $Ec$ as the polymer approaches full extension \citep{hinch2021oldroyd,zinelis2023transition}. Various extensions of the Oldroyd-B equations have been developed to account for such nonlinearity, either in equations~\eqref{sigmap} and~\eqref{Aupperconv} or in the solvent contribution in equation~\eqref{taus} \citep{de1974coil,tanner2000engineering,mckinley2002filament,alves2021numerical}. In this study, we employ the Oldroyd-B model to include the two primary effects of the polymer addition: the additional stress ($Ec$) and polymeric liquid memory ($De$) \citep{snoeijer2020relationship}. Our aim is to provide a comprehensive understanding of the entire $Ec$-$De$ parameter space (figure~\ref{schematic}b).
\oo 
However, it is crucial to note that the Oldroyd-B model, while serving as a useful baseline, cannot accurately reproduce the finite-time breakup of viscoelastic filaments \citep{eggers2020self} or the full complexity of interface rupture \citep{lohse-2020-pnas}. These limitations warrant caution when interpreting the final stages of jet thinning and droplet formation, particularly in scenarios involving strong polymer stretching.
\bb

\subsection {Methods}\label{sec:methods}

We employ the open-source software Basilisk C \citep{basilliskpopinet, popinet2015quadtree} to solve the governing equations outlined in \S~\ref{subsec:governingEqn}.
\oo
To solve the Oldroyd-B viscoelastic constitutive relation (equation~\eqref{oldroydb}), Basilisk C uses the log-conformation method \citep{fattal2004constitutive} implemented by \citet{lopez2019adaptive} which has been used extensively at finite $De$ \citep{turkoz2018axisymmetric, turkoz2021simulation}. To explore the entire $Ec$-$De$ parameter space  (figure~\ref{schematic}c), we have extended the log-conformation formulation to solve equations~\eqref{sigmap} and~\eqref{Aupperconv}. In the spirit of Basilisk C, this code is detailed open-source at \citet{vatsalElastoFlow2024}.\bb\,
The rest of the governing equations are solved using the one-fluid approximation \citep{tryggvason2011direct}, with surface tension incorporated as singular body force at \DL{the} liquid-gas interface \citep{brackbill1992continuum}.
To account for the gas phase, in addition to the dimensionless parameters described in \S~\ref{sec:Intro} and \S~\ref{subsec:governingEqn}, we maintain constant density and viscosity ratios of $\rho_{r} = \rho_{g}/\rho_{s} = 10^{-3}$ and $\eta_{r} = \eta_{g}/\eta_{s} = 2 \times 10^{-2}$, respectively.
The liquid-gas interface is tracked using the volume of fluid (VoF) method, governed by the advection equation

\begin{align}
	\frac{\partial \Psi}{\partial t} + \boldsymbol{\nabla\cdot}\left(\Psi\boldsymbol{u}\right) = 0,
	\label{volfracconserve}
\end{align}

\noindent where $\Psi$ represents the VoF color function. We implement a geometric VoF approach, reconstructing the interface at each timestep and applying surface tension forces as singular forces \citep{popinet2009accurate, brackbill1992continuum}

\begin{align}
	\boldsymbol{f}{\gamma} \approx \kappa \boldsymbol{\nabla} \Psi,
	\label{f}
\end{align}

\noindent with curvature $\kappa$ calculated using the height-function method \citep{popinet2018numerical}. The explicit treatment of surface tension imposes a time step constraint based on the smallest capillary wave oscillation period \citep{popinet2009accurate}. Yet another time step restriction, usually more relaxed than the surface tension one, comes from the explicit treatment of the polymeric stress term $\boldsymbol{\sigma}_p$.
We impose no-penetration and free-slip conditions at wall boundaries to avoid wall-shear effects, with outflow conditions at the top boundary to prevent droplet rebound. Pressure gradients are set to zero at domain boundaries for both liquid and gas phases.

The initial bubble shape is determined by solving the Young-Laplace equations for quasi-static equilibrium \citep{princen1963shape,toba1959drop,villermaux2022bubbles,VatsalThesis}.
While the shape's asymmetry increases with the Bond number $Bo$, we focus on the limit $Bo \to 0$, setting $Bo = 0.001$ to regularize the singularity at the sphere-plane intersection. This results in a near-spherical initial cavity shape (figure~\ref{schematic}a-i).
\oo We stress that here we assume that the bubble has resided at the liquid-gas interface for a duration far exceeding the polymeric medium's relaxation time, ensuring that elasticity does not influence the initial configuration \citep{ari2024bursting}.\bb\,
During the bubble cap bursting, the film cap retracts almost instantaneously \oo(once again, we neglect the influence of elasticity)\bb, after which the capillary waves are generated.
As we are interested only in the bubbe cavity collapse, the simulations begin with an open cavity without the thin cap (figure~\ref{schematic}a-ii), as also done similarly in recent studies \citep{deike2018dynamics, gordillo2019capillary,sanjay2021bursting}. The computational domain spans $8R_0 \times 8R_0$, discretized using quadtree grids with adaptive mesh refinement (AMR) \citep{popinet2009accurate}. Error tolerances for the VoF color function, curvature, velocity, and order parameter $\boldsymbol{\mathcal{A}}$ are set to $10^{-3}$, $10^{-6}$, $10^{-3}$, and $10^{-3}$, respectively.

In this work, following our earlier study \citep{sanjay2021bursting}, most simulations maintain a minimum grid size of $\Delta = R_0/512$, which dictates that, to get consistent results, 512 cells are required across the bubble radius while using uniform grids.
We have also used an increased resolution ($\Delta = R_0/1024$ for high $De$ cases and $\Delta = R_0/2048$ near transitions) as needed. 
\oo
These resolutions are consistent with previous studies by \citet{berny2020role,berny2021statistics} on bubble bursting and \citet{turkoz2018axisymmetric,turkoz2021simulation} on visco-elastic thinning with a maximum level of resolution of 14 (for $\Delta = R_0/2048$ and domain size $L_0 = 8R_0$).
\bb 
We have carried out extensive grid independence studies to ensure that changing the grid size does not influence the results \oo (see appendix~\ref{app:gis})\bb. We refer the readers to \citet{popinet2015quadtree,VatsalThesis,Sanjay2024code} for further details of the numerical method used in this work.

\begin{figure}
	\includegraphics[width=\textwidth]{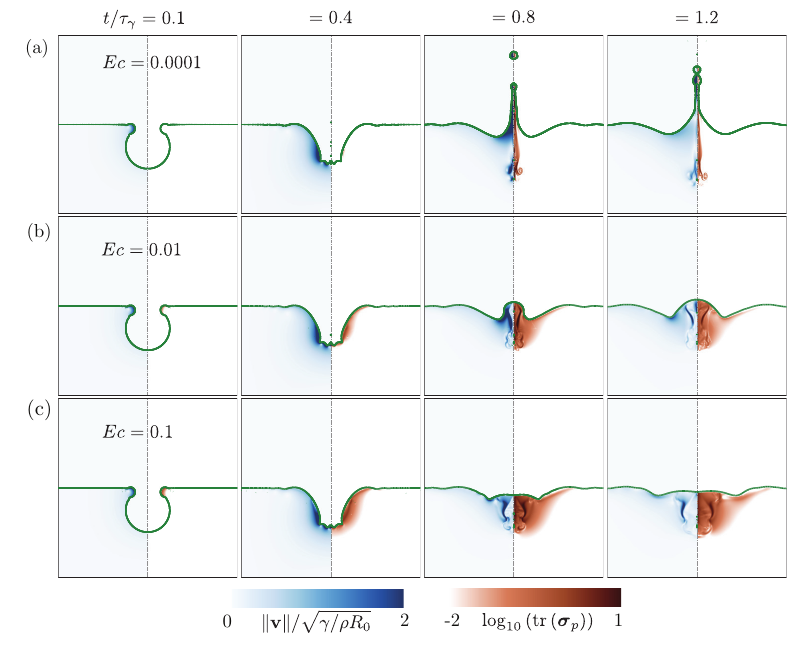}
	\caption{Temporal evolution of \DL{the} bubble cavity collapse at $De \to \infty$ and $Oh_s = 0.025$ for $Ec =$ (a) $0.0001$, (b) $0.01$\AKD{,} and (c) $0.1$. \AO{The color scheme in the left panel of each snapshot} represents the magnitude of the velocity field normalized by the inertiocapillary velocity, \AO{while on the right panel of each snapshot, it} shows the trace of the elastic stress $\boldsymbol{\sigma}_p$ that represents twice the elastic energy stored in polymeric deformations on a $\log_{10}$ scale. See also the supplementary movies SM1.}
	\label{facets_time_highDe}
\end{figure}

\section{Influence of polymers}
\label{sec:polymers}

This section phenomenologically describes the influence of polymers on the bursting bubble process by investigating how varying the elastocapillary number $Ec$ influences the formation of Worthington jets and droplet ejection. We focus on two limiting cases: \DL{polymeric solutions} with permanent memory exhibiting affine motion ($De \to \infty$) and those with poor memory ($De \to 0$).

\subsection{Polymeric liquids with permanent memory}
\label{sec:LargeDe}

We begin our analysis by considering the limit of $De \to \infty$, where the \DL{polymeric solutions} feature affine motion (equation~\eqref{eq:affineMotion}) and \VS{maintain a permanent memory of their initial condition and subsequent deformations} without relaxation during the process timescale.
Figure~\ref{facets_time_highDe} illustrates representative cases in viscoelastic media for $Oh_s = 0.025$ and varying elastocapillary numbers ($Ec$). The figure presents a temporal evolution of the interface profile (green line) alongside \DL{with the} velocity magnitude on the left and the trace of elastic stress $\boldsymbol{\sigma}_p$ on the right.
Remarkably, despite all cases exhibiting a total Ohnesorge number of infinity ($Oh_s + Oh_p \to \infty$), which typically implies highly viscous behavior (see figure~\ref{facets_time_Oh_Newt}), low $Ec$ scenarios demonstrate dynamics qualitatively resembling Newtonian fluids. In these cases, capillary waves drive the collapse of a bubble cavity, converging at its bottom to form a Worthington jet that subsequently fragments into droplets (see figure~\ref{facets_time_highDe}a). Intuitively, the elastic stresses are concentrated near the axis of symmetry where the strain is maximum \citep{turkoz2018axisymmetric,eggers2020self}.
The process concludes within a finite timescale ($\sim \tau_\gamma$), resulting in a regular limit as $Ec \to 0$. As a result, the system's behavior deviates gradually from the Newtonian case at $Ec = 0$, exhibiting a continuous transition as \DL{the} elasticity increases.
This absence of singularity contrasts with elastic Taylor--Culick-type retractions, where an infinite process timescale allows the elastic stresses to develop\DL{,} leading to distinct behaviors for $Ec = 0$ and $Ec \to 0$ \citep{BertinSanjay2024}\DL{, i.e., a singular limit.}

We stress that in this limit, the jet breakup occurs due to finite grid resolution in our numerical code \citep{lohse-2020-pnas,chirco2022manifold,kant2023bag}. We cannot differentiate between a case of drop detach\DL{ment} from the jet or \DL{the case when} they are still connected through a thin filament--also known as the beads-on-a-string structure \citep{hosokawa2023phase, clasen2006beads, pandey2021elastic, zinelis2023transition}. 
Although current simulations fully resolve other aspects, they \DL{cannot} resolve these finest threads, which may have subgrid cell sizes depending on the $Ec$.
At higher grid resolutions, we expect to recover the beads-on-a-string configuration, as the Oldroyd-B model does not yield a finite time breakup singularity in the infinite $De$ regime, instead converging to a finite filament \citep{eggers2020self,turkoz2018axisymmetric,turkoz2021simulation}.
To prevent infinite thread thinning, a nonlinear elastic model could also be employed (see \S~\ref{subsec:governingEqn} for further discussions).

\begin{figure}
	\includegraphics[width=\textwidth]{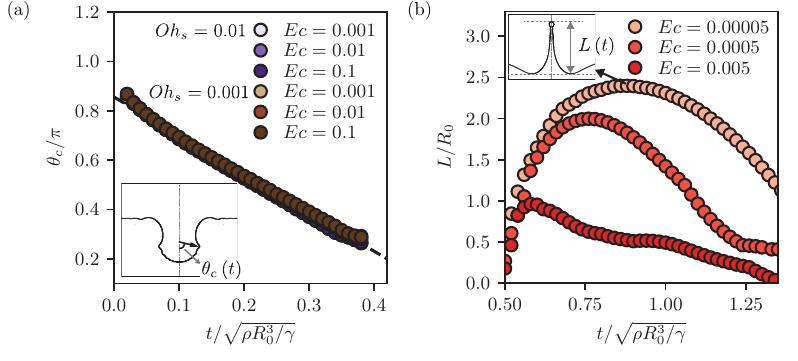}
	\caption{(a) Trajectory of the maximum curvature capillary wave parameterized using the angle $\theta_c(t)$ as depicted in the inset at $De \to \infty$ for different $Oh_s$ and $Ec$. (b) Evolution of the jet length $L(t)$ at $Oh_s = 0.04$ and $De \to \infty$ for different $Ec$.}
	\label{theta_time}
\end{figure}

As $Ec$ increases, we observe jet formation without droplet ejection (figure~\ref{facets_time_highDe}b). At higher $Ec$ values, even jet formation is suppressed due to elevated elastic resistance (figure~\ref{facets_time_highDe}c).
Notably, while polymeric effects significantly influence the dynamics after the convergence of capillary waves (figure~\ref{facets_time_highDe}, $t/\tau_\gamma = 0.8, 1.2$), the propagation of capillary waves (figure~\ref{facets_time_highDe}, $t/\tau_\gamma = 0.1, 0.4$) remains largely unaffected. Figure~\ref{theta_time}(a) quantifies the trajectories of these capillary waves across three orders of magnitude variation in $Ec$ at two different $Oh_s$. The capillary wave speed is independent of both liquid and polymeric control parameters, mirroring the behavior observed in Newtonian media \citep{gordillo2019capillary} and contrasting those for viscoplastic media \citep{sanjay-2022-JFM}. \rev{The independence of capillary wave speed on the polymeric control parameters has also been reported in the experiments \citep{cabalganteeffect}.}
Following capillary wave collapse, the Worthington jet initially elongates to a maximum length ($L_{\text{max}}$) before retracting. As shown in Figure~\ref{theta_time}(b) for $Oh_s = 0.04$, $L_{\text{max}}$ decreases with increasing $Ec$ due to stronger resistive stresses.

\begin{figure}
	\includegraphics[width=\textwidth]{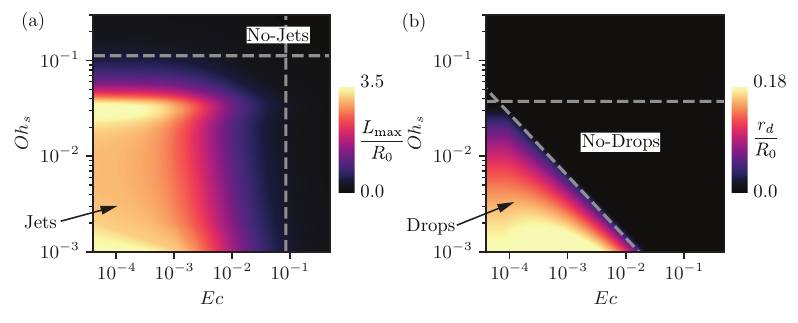}
	\caption{(a) The maximum jet length $L_{\text{max}}$ at $De \to \infty$ in the $Ec$-$Oh_s$ phase space\AKD{, depicted by the colormap, where the lighter region corresponds to higher values}. For the Newtonian liquid \DL{$\left(Ec \to 0\right)$}, the \VS{jetting transition} occurs at $Oh_s = 0.11$, denoted by the horizontal dotted line. Due to the elastic effect\AO{s}, this transition occurs at $Ec = 0.086$, as depicted by the vertical dotted line. (b) The size of the first droplet at $De \to \infty$ in the $Ec$-$Oh_s$ phase space. For the Newtonian liquid, \VS{the dropping transition} is observed at $Oh_s = 0.0375$, denoted by the horizontal dotted line. Further, the transition due to elastic effects is very sensitive to $Oh_s$ and is shown by the inclined dotted line.}
	\label{highDe}
\end{figure}

Figure~\ref{highDe}(a) presents a phase map of $L_{\text{max}}$, compiled from approximately 100 simulations.
For Newtonian liquids, $L_{\text{max}}$ peaks near $Oh_s \approx 0.03$, corresponding \AO{to the value of} observed hydrodynamic singularities \citep{zeff2000singularity,lohse2003bubble,eggers2015singularities,yang2020multitude}, before decreasing at higher $Oh_s$ \citep{duchemin2002jet,deike2018dynamics,gordillo2019capillary}.
Jet formation ceases altogether beyond a critical value of $Oh_c = 0.11$ \citep{sanjay2021bursting} (defined here when $L_\text{max} < 0.3R_0$).
As $Ec$ increases, viscoelastic effects become significant. $L_{\text{max}}$ decreases monotonically with $Ec$ due to increased elastic resistance, with jet formation suppressed beyond $Ec=0.086$. Unlike the non-monotonic relationship between $L_{\text{max}}$ and $Oh_s$, where increasing $Oh_s$ initially produces thinner and faster jets, the $L_{\text{max}}(Ec)$ relationship remains consistently monotonic. Even the $Oh_s$-sensitive singular Worthington jets disappear with increasing $Ec$. Notably, the critical $Ec$ values for these transitions appear \DL{to be} largely independent of $Oh_s$, in contrast to the $Oh_s$-dependent behavior observed in the Newtonian limit.

The emerging Worthington jet may eject multiple droplets. For Newtonian liquids, \oo predictions\bb\, for the first droplets' size $r_d$ are well understood (see appendix~\ref{app:newtonian_limit} and \cite{ganan2017revision,blanco2020sea}). $r_d$ decreases with $Oh_s$ until $Oh_s \approx 0.0375$, beyond which \DL{the} droplet breaks \DL{from the jet} due to the Rayleigh--Plateau instability and \AKD{falls} downwards. Our analysis focuses on droplets propagating away from the source, excluding those with downward velocity upon breakup (observed in Newtonian media for $0.0375 < Oh_s < 0.045$). For elastic cases, despite unresolved filaments connecting droplets and jets, we have rigorously verified the convergence of the first droplet's size to at least 10\% accuracy.
Figure~\ref{highDe}(b) illustrates a phase map of the first droplet's size $r_d$, revealing intriguing differences from the jet behavior. While $r_d$ follows the same trend with $Oh_s$ \oo observed at Newtonian limits\bb\, and remains invariant of $Ec$ below critical values, the critical $Ec$ for droplet suppression differs from that of jet suppression. As the jet width is determined solely by $Oh_s$, \KZ{independently} of $Ec$, the first emerging droplet's size also remains independent initially. However, as $Ec$ increases further, rising elastic stresses suppress droplet formation more \AO{abruptly} than jet formation.
The critical values $Ec_d$ for \VS{the transition between jet formation with and without droplet breakup (dropping transition)} are sensitive to $Oh_s$, with the critical $Ec_d$ decreasing as $Oh_s$ increases. This trend \AO{is in stark contrast} with \VS{the transition from jet formation to jet suppression (jetting transition)}, where critical $Ec$ values remain largely $Oh_s$-independent.

\begin{figure}
	\centering
	\includegraphics[width=\textwidth]{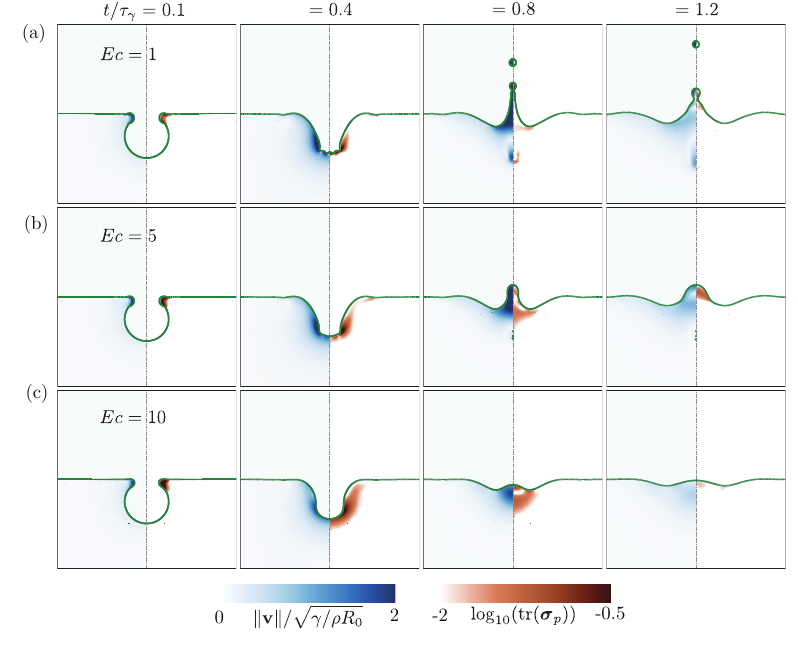}
	\caption{Temporal evolution of bubble cavity collapse at $De = 0.01$ and $Oh_s = 0.025$ for $Ec =$ (a) $1$, (b) $5$, and (c) $10$. \AO{The color scheme in the left panel of each snapshot} represents the magnitude of the velocity field normalized by the inertiocapillary velocity, \AO{while on the right panel of each snapshot, it shows the trace} of the elastic stress $\boldsymbol{\sigma}_p$ that represents twice the elastic energy stored in polymeric deformations on a $\log_{10}$ scale. See also the supplementary movies SM2.}
	\label{factes-LowDe}
\end{figure}

\begin{figure}
	\centering
	\includegraphics[width=\textwidth]{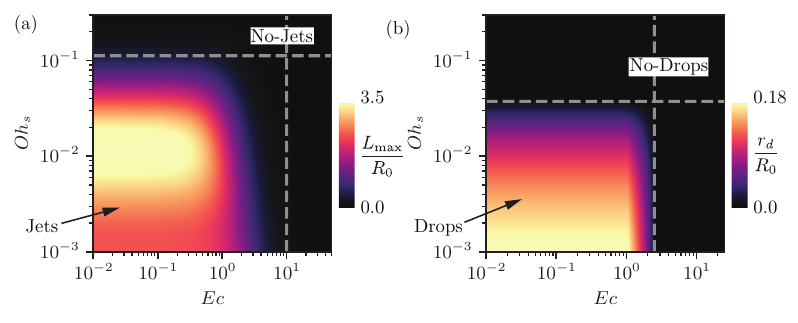}
	\caption{(a) The maximum jet length $L_{\text{max}}$ at $De = 0.01$ in the $Ec$-$Oh_s$ phase space\AKD{, depicted by the colormap, where the lighter region corresponds to higher values}. For the Newtonian liquid, the \VS{jetting transition} occurs at $Oh_s = 0.11$, denoted by the horizontal dotted line. Due to the elastic effects, this transition occurs at $Ec = 9.3$, as depicted by the vertical dotted line. (b) The size of the first droplet at $De = 0.01$ in the $Ec$-$Oh_s$ phase space. For the Newtonian liquid, \VS{the dropping transition} is observed at $Oh_s = 0.0375$, denoted by the horizontal dotted line. Further, the $Oh_s$-independent transition due to elastic effects occurs at $Ec= 2.5$, as shown by the vertical dotted line.}
	\label{smallDe}
\end{figure}

\subsection{Polymeric liquids with poor memory}
\label{sec:shortmemory}

This section examines the dynamics in media with \VS{a poor memory of its initial conditions and subsequent deformations. ($De \to 0$)}.
For sufficiently small Deborah numbers $De$, \DL{the} polymers relax rapidly, resulting in elastic stresses \DL{of the polymeric liquid} that are considerably lower than those observed in cases where $De \to \infty$. The stress relaxation also results in the dissipation of elastic energy stored \AO{in stretched polymers}. Figure~\ref{factes-LowDe} illustrates representative cases for $De = 0.01$, showcasing three distinct regimes as a function of the elastocapillary number ($Ec$). The figure presents a temporal evolution of the interface profile (green line) alongside velocity magnitude on the left and the trace of elastic stress $\boldsymbol{\sigma}_p$ on the right for $Ec = 1$, $5$, and $10$. For $Ec = 1$ (figure~\ref{factes-LowDe}a), we observe a slender Worthington jet that forms a droplet.
As $Ec$ increases to $5$ (figure~\ref{factes-LowDe}b), the jet persists but fails to produce a droplet.
At $Ec = 10$ (figure~\ref{factes-LowDe}c), jet formation is completely suppressed, with the interface showing only slight deformations during cavity relaxation.
The qualitative trends with respect to the elastocapillary number ($Ec$) remain consistent as compared to those in \S~\ref{sec:LargeDe}. However, the critical $Ec$ values for different regimes differ markedly from those observed at $De \to \infty$. Notably, jet formation and droplet production persist at $Ec = 1$ (figure~\ref{factes-LowDe}a), despite this value being an order of magnitude higher than the critical $Ec$ for the \VS{jetting} transition at infinite $De$. \AO{This difference underscores the dependence of transition thresholds on $De$.}

\AO{To further interpret the jetting dynamics and drop formation}, figure~\ref{smallDe} presents phase maps \VS{illustrating} the behavior of maximum jet lengths ($L_{\text{max}}$) and first droplet sizes ($r_d$) for $De = 0.01$. Figure~\ref{smallDe}(a) \DL{shows} $L_{\text{max}}$ across a range of $Ec$ and $Oh_s$ values.
For low $Ec$, $L_{\text{max}}$ shows Newtonian-like $Oh_s$ dependence. As $Ec$ increases, $L_{\text{max}}$ decreases monotonically until jet formation ceases beyond an $Oh_s$-independent critical $Ec_j$, mirroring the infinite $De$ limit behavior.
Figure~\ref{smallDe}(b) maps the $r_d$, showing $Ec$-independent droplet sizes \oo that are equal to values at the Newtonian limit\bb, until near the transition point, where droplet formation is suppressed. For $De \ll 1$, the critical $Ec_d$ for the \VS{dropping} transition exhibits minimal $Oh_s$-dependence, contrasting with the $Oh_s$-sensitive behavior at infinite $De$.
Comparing these results to the $De \to \infty$ limit reveals persistent fundamental regimes across different $De$ values, but \DL{the} transition thresholds are highly sensitive to \DL{the polymeric liquid's} relaxation time. Critical $Ec$ values for both jet and droplet suppression are significantly higher at low $De$ compared to the infinite $De$ limit, indicating that rapid \DL{relaxation of polymeric stresses} allows jet and droplet formation at higher $Ec$ values. This low $De$ behavior suggests an interplay between elastic and viscous effects, explored further in \S~\ref{sec:regimes}.

\section{Regime Map}
\label{sec:regimes}

% \vsy{\#TODO-DONE: following reviewer 3, I have basically rewritten \S~4.1 and 4.2.2. @Ayush: please check that I did not break any logic.}

\oo
The bursting bubble dynamics in viscoelastic media exhibit distinct behavior compared to Newtonian fluids.
Our analysis reveals three well-defined regimes: (i) jets that form droplets, (ii) jets without droplet formation, and (iii) complete suppression of jets. While viscoelasticity significantly modifies jet dynamics, the capillary wave propagation prior to jet formation remains remarkably unaffected.
This section explores the transitions between these regimes across the $Ec$-$De$ phase space, extending our earlier analysis of the limiting cases $De \to \infty$ and $De \to 0$ from \S~\ref{sec:polymers}.
\bb

\subsection{Summary of the different regimes}
\label{subsec:summaryRegimes}

\begin{figure}
	\centering
	\includegraphics[width=\textwidth]{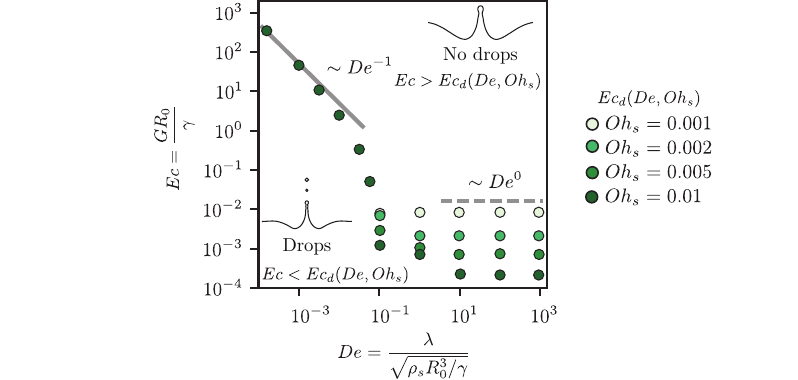}
	\caption{The elastocapillary-Deborah number ($Ec$-$De$) phase map delineating the transition between the regimes: (i) jets forming droplets and (ii) jets without droplet formation. The data points represent the critical elastocapillary number $Ec_d(De, Oh_s)$ at which this transition occurs. The transition behavior exhibits distinct characteristics in different limits: as $De \to \infty$, the transition occurs at a constant $Ec$ which is highly sensitive to $Oh_s$ (see the gray dashed line showing $Ec_d \sim De^0$), while for $De \to 0$, the transition is $Oh_s$-independent and occurs at constant $Oh_p$ (see the gray solid line showing $Ec_d \sim De^{-1}$, i.e., $Oh_{p,d} \sim De^0$).}
	\label{fig:transitionDrops}
\end{figure}

\begin{figure}
	\centering
	\includegraphics[width=\textwidth]{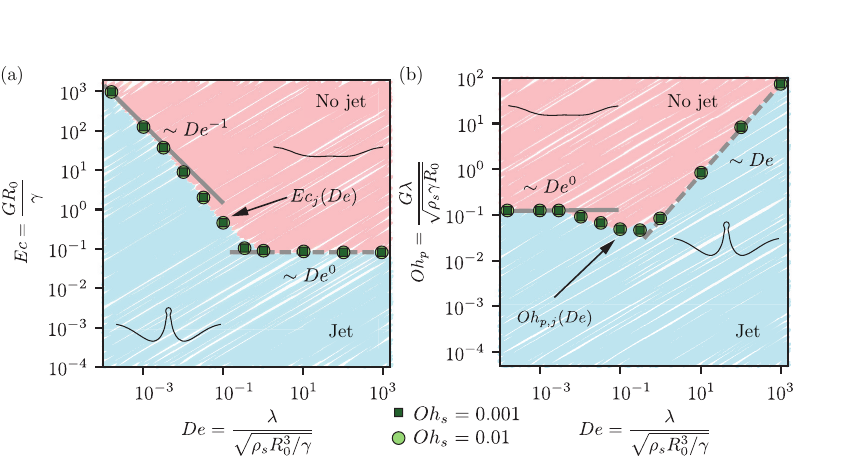}
	\caption{(a) The elastocapillary-Deborah number ($Ec$-$De$) and (b) the polymeric Ohnesorge-Deborah number ($Oh_p$-$De$) phase map delineating the transition between the regimes: (ii) jets without droplet formation and (iii) absence of jet formation. The data points represent the $Oh_s$-independent critical elastocapillary number \AO{$Ec_j(De)$} at which this transition occurs. The transition behavior exhibits distinct characteristics in different limits: as $De \to \infty$, \AKD{the} transition occurs at a constant $Ec$ (see gray dashed line showing $Ec_d \sim De^0$), while for $De \to 0$, \AKD{the} transition occurs at constant $Oh_p$ (see gray solid line showing $Ec_d \sim De^{-1}$, i.e., $Oh_{p,d} \sim De^0$).}
	\label{fig:transitionJets}
\end{figure}

\oo
The transitions between these regimes depend on both $Ec$ and $De$, exhibiting markedly different characteristics in two limiting cases: $De \to \infty$ and $De \to 0$.
Figure~\ref{fig:transitionDrops} maps these transitions in the elastocapillary-Deborah number ($Ec$-$De$) phase space, delineating the boundaries between droplet-forming jets and jets without droplets. Figure~\ref{fig:transitionJets} complements this by illustrating the transition to complete jet suppression. Notably, the infinite $De$ asymptotic behavior extends down to $De \approx 1$, reflecting that polymers lack sufficient time to relax when relaxation times exceed the process timescale.
\bb

For polymeric liquids with long relaxation times ($De \gg 1$), we observe that:

\begin{enumerate}
	\item the dropping transition occurs at $Ec_d(Oh_s)$, with strong $Oh_s$ dependence (Figures~\ref{highDe}b and \ref{fig:transitionDrops}), and
	\item the jetting transition occurs at $Ec_j \approx 0.086$, independent of $Oh_s$ (Figure~\ref{fig:transitionJets}a).
\end{enumerate}

Conversely, for polymeric liquids with short relaxation times ($De \ll 1$), we find that both transitions are $Oh_s$-independent and occur at constant polymeric Ohnesorge number $Oh_p = Ec \times De$:

\begin{enumerate}
	\item the dropping transition occurs at $Oh_{p,d} \approx 0.048$ (Figure~\ref{fig:transitionDrops}) and
	\item the jetting transition occurs at $Oh_{p,j} \approx 0.129$ (Figure~\ref{fig:transitionJets}b).
\end{enumerate}

\oo

These behaviors reflect fundamentally different physical mechanisms: at high $De$, depending on $Oh_s$, the medium behaves like an elastic solid ($Oh_s \to 0$) or Kelvin-Voigt solid (finite $Oh_s$). However, at low $De$, polymer addition manifests as an enhanced viscous effect characterized by $Oh_p$. \bb \rev{The trend of dropping transition in small $De$ regime is qualitatively similar to recently reported experimental observation \citep{cabalganteeffect}. Although, a quantitative comparison cannot be made due to significant differences in $Bo$. }
We further investigate the jetting transition using slender jet equations in \S~\ref{app:elasticeffects} following similar approaches by \citet{driessen2013stability,gordillo2020impulsive,zinelis2023transition,sen2024elastocapillary}.

\begin{figure}
	\centering
	\includegraphics[width=\textwidth]{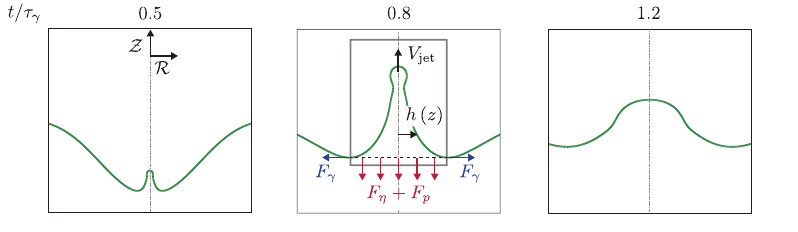}
	\caption{
		Temporal evolution of the Worthington jet for a representative case, where the jet emerges, reaches a maximum, and is pulled to merge with the liquid bath. The control volume contains the jet region, as shown by the region within the gray lines. Here, $h(z,t)$ is the width of the jet, which becomes $h_{\text{base}}$ at the base of the jet. The capillary force at the jet base is $F_\gamma = \gamma \left(2 \pi h_{\text{base}} \right)$ that acts radially outwards. At the same time, the elastic and viscous stresses act at the base of the jet as $\DL{F_\eta + F_p  = \left( \sigma_{\eta, \text{base}} + \sigma_{p, \text{base}}\right) \pi h_{\text{base}}^2}$.
	}
	\label{fig:jetSchematic}
\end{figure}

\subsection{What sets the different transitions, and what do we learn from these transitions?}
\label{app:elasticeffects}

To understand the mechanisms governing bubble cavity collapse, we analyze jet dynamics using a control volume approach (figure~\ref{fig:jetSchematic}). Employing the slender jet approximation \citep{shi1994cascade, eggers2015singularities,driessen2013stability}, given the small radial-to-axial length scale ratio, the vertical momentum equation for the jet reads

\begin{align}
	\rho_s \left(	\frac{\partial v}{\partial t} + v\frac{\partial v}{\partial z} \right) = - \gamma \frac{\partial \kappa} {\partial z} + \frac{1}{h^2} \frac{\partial}{\partial z}\left[ h^2 \left( 3\eta_s \frac{\partial v}{\partial z} + G\left(\mathcal{A}_{zz}-1\right) \right) \right].
	\label{slenderjet}
\end{align}

\noindent Here, $v(z,t)$ is the radially averaged jet velocity, and the shape of this jet is $h(z,t)$. \AO{We define a control volume containing the emerging jet that is always bounded by the inflection points at the interfaces, see figure \ref{fig:jetSchematic}b}. Integrating over \AO{this} control volume (with differential volume element $d\Omega = \pi h(z,t)^2\mathrm{d}z$) yields the force balance \citep{trouton1906coefficient}:

\begin{align}
	\frac{d \mathcal{M}_{\text{jet}}}{d t} = 3 \eta_sh^2\frac{\partial v}{\partial z}\Bigg|_{\text{base}} + Gh^2(\mathcal{A}_{zz}-1)\Bigg|_{\text{base}} = \AO{\left(\sigma_{\eta,\text{base}} + \sigma_{p,\text{base}}\right)\pi h_{\text{base}}^2}
	\label{forcebalance}
\end{align}

\noindent where
% $\Pi_{\text{jet}}(t) = \int_{\Omega(t)}\rho_sv(z,t)^2\pi h(z,t)^2\mathrm{d}z$
\AKD{$\mathcal{M}_{\text{jet}}(t) = \int_{\Omega(t)}\rho_sv(z,t)\pi h(z,t)^2\mathrm{d}z$}
denotes the momentum of the jet.
The capillary stress (first term on the right-hand side of equation~\eqref{slenderjet}) integral vanishes due to orthogonal interface intersection with the control volume (see \citet{marchand2011surface} and p.~16-21, \citet{munro2019coalescence}). \AO{We chose this control volume because of its vanishing integral feature.}
Furthermore, the integral of the second term on the right-hand side forms an exact integral which vanishes at the tip where it is zero owing to $h(z = L_{\text{max}}(t)) = 0$.
Consequently, jet evolution depends solely on stresses at the base: viscous ($\sigma_{\eta,\text{base}}(t)$) and elastic ($\sigma_{p,\text{base}}(t)$).
For relevant $Oh_s$ values, $\sigma_{\eta,\text{base}}(t)$ is too weak to suppress the Worthington jet. Numerical simulations allow \AO{us} \DL{to} estimate $\sigma_{p,\text{base}}(t)$. As \DL{the} capillary waves collapse, \DL{the} base elastic stress reaches a global maximum\DL{,} before decreasing \DL{again at later times}. Jet formation occurs if inertial flow focusing is sufficiently strong at \DL{the} peak elastic stress. We will now evaluate this competition for the two limits of $De$.

\begin{figure}
	\centering
	\includegraphics[width=\textwidth]{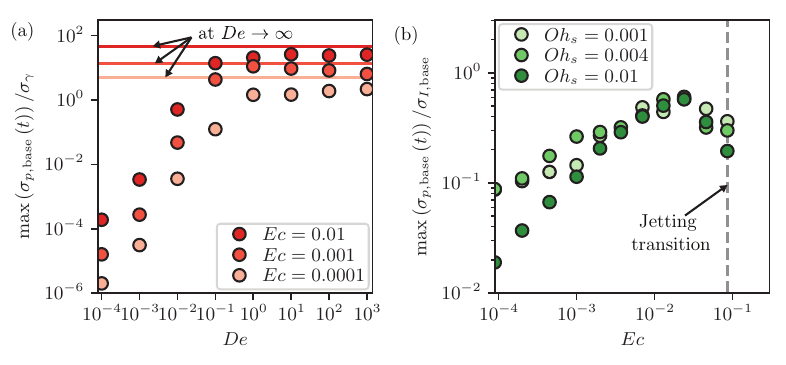}
	\caption{(a) Evolution of the maximum elastic stress at the jet base ($\text{max}\left(\sigma_{p,\text{base}}(t)\right)$), normalized by the Laplace pressure scale $\sigma_\gamma = \gamma/R_0$, as a function of $De$ for different $Ec$ at $Oh_s = 0.001$. {\oo Note that $Oh_p = Ec\times De$. \bb} (b)  Comparison of the resistive elastic stress $\text{max}\left(\sigma_{p,\text{base}}(t)\right)$ in the high $De$ regime $\left(\to \infty \right)$ against the inertial stresses $\sigma_{I,\text{base}}$, plotted against $Ec$ for different $Oh_s$.}
	\label{stress_De}
\end{figure}

\subsubsection{The limit of $De \to \infty$}

Figure~\ref{stress_De}(a) shows that for $De > 1$, the maximum elastic stress $\text{max}\left(\sigma_{p,\text{base}}(t)\right)$ reaches a plateau, dependent only on $Ec$. This $De$-independence coincides with the extent of infinite $De$ asymptotes featured in the transitions discussed in \S~\ref{subsec:summaryRegimes}.
The upper limit of elastic resistance competes with inertial flow focusing to inhibit jet formation. We quantify the inertial stresses at peak elastic stress using:

\begin{align}
	\sigma_{I,\text{base}} = \frac{2}{h_{\text{base}}^2}\int_{o}^{h_{\text{base}}} \rho_s v^2 h\mathrm{d}h,
\end{align}

\noindent where $h_{\text{base}}$ is the jet width at its base (see figure~\ref{fig:jetSchematic}).
Figure~\ref{stress_De}(b) reveals that the ratio of elastic to inertial stresses is largely independent of $Oh_s$. As $Ec$ increases, this ratio reaches a maximum beyond which jet suppression occurs.
It is important to note that the apparent decrease in \DL{this stress ratio with increasing $Ec$ and $Oh_s$} near the \VS{jetting} transition in figure~\ref{stress_De}(b) \DL{occurs due to a decrease in both the polymeric and inertial stresses in this region of the parameter space.}

\begin{figure}
	\centering
	\includegraphics[width=\textwidth]{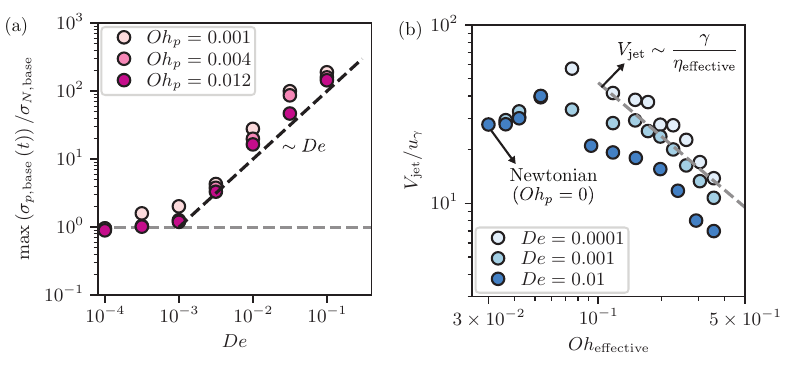}
	\caption{(a) Evolution of the maximum elastic stress at the jet base ($\text{max}\left(\sigma_{p,\text{base}}(t)\right)$), normalized by the Newtonian-like viscous stress $\sigma_{N,\text{base}}$ with viscosity $\eta_{p} = G\lambda$, as a function of $De$ for different $Oh_p$ at $Oh_s = 0.001$. The gray dashed horizontal line represents $\text{max}\left(\sigma_{p,\text{base}}(t)\right) \approx \sigma_{N,\text{base}}$ while the black dashed line serves as a guide to the eye representing $\text{max}\left(\sigma_{p,\text{base}}(t)\right)/\sigma_{N,\text{base}} \sim De$. {\oo Note that $Ec = Oh_p/De$. \bb} (b) The variation of jet's tip velocity $V_{\text{jet}}$, normalized by the inertiocapillary velocity $u_\gamma = \sqrt{\gamma/\rho_sR_0}$, with $Oh_{\text{effective}} = 3Oh_s + 2 Oh_p$ at different $De$ and $Oh_s = 0.01$. The gray dashed line represents $V_{\text{jet}} \sim \gamma/\eta_{\text{effective}}$.}
	\label{stressratio_De}
\end{figure}

\subsubsection{The limit of $De \to 0$}

\oo
In the zero $De$ limit, polymeric liquids exhibit additional viscous effects characterized by the polymeric Ohnesorge number $Oh_p$ (also see \S~\ref{subsec:summaryRegimes}).
The maximum elastic stress $\text{max}(\sigma_{p,\text{base}}(t))$, when normalized by the Newtonian-like viscous stress $\sigma_{N,\text{base}}$, collapses for all $Oh_p$ as $De \to 0$, where
\bb
%
%To quantify this stress,
%we evaluate the maximum elastic stress $\text{max}(\sigma_{p,\text{base}}(t))$ normalized by a Newtonian-like viscous stress $\sigma_{N,\text{base}}$, calculated when the elastic stress reaches its maximum using

\begin{align}
	\sigma_{N,\text{base}} =  \frac{2}{h_{\text{base}}^2}\int_{o}^{h_{\text{base}}} G\lambda\frac{\partial v}{\partial z} h\mathrm{d}h.
\end{align}

%\noindent Here, $\eta_p = G\lambda$ \DL{represents the polymeric contribution to the shear viscosity of the polymeric liquid.}
\oo
As $De$ approaches unity, marking the onset of the infinite $De$ asymptotic regime, the elastic stress scales as $\text{max}(\sigma_{p,\text{base}}(t)) \sim De \times \sigma_{N,\text{base}}$.
%we observe a scaling behavior of $\text{max}(\sigma_{p,\text{base}}(t)) \sim De \times \sigma_{N,\text{base}}$.
This scaling remarkably resembles that predicted by \citet{boyko2024flow} for flow in a slowly varying contraction at the infinite $De$ asymptote, despite significant geometric differences. While our study focuses on free surface flows and \citet{boyko2024flow} examined contraction geometries, this unexpected similarity hints at a potentially universal behavior near the infinite $De$ asymptote. To further \VS{examine} this intriguing connection, a similar closed-form $De$ expansion for free surface flows is necessary. However, we caution that this scaling approach to the infinite $De$ asymptote could be system-dependent \citep{hinch2024fast}.

At zero $De$, the elastic stress reduces to a Newtonian-like viscous stress with polymeric viscosity $\eta_p$, yielding $\sigma_p \approx 2G\lambda\boldsymbol{\mathcal{D}}$ for Oldroyd-B rheology. The force balance in equation~\eqref{forcebalance} becomes
\bb
%Here, we focus at the zero $De$ limit, where the elastic stress can be replaced with a Newtonian-like viscous stress with polymeric viscosity $\eta_p$ such that the \VS{jetting} transition occurs at constant polymeric $Oh_p$.
%The polymeric stress contribution becomes $\sigma_p \approx 2G\lambda\boldsymbol{\mathcal{D}}$ (for the Oldroyd-B rheology, see also equation~\eqref{oldroydb}).
%Consequently, we can rewrite the force balance in equation~\eqref{forcebalance} as

\begin{align}
	\frac{d \mathcal{M}_\text{jet}}{d t} = \left(3\eta_s + 2G\lambda\right)h^2\frac{\partial v}{\partial z}\Bigg|{\text{base}},
	\label{etaeffect}
\end{align}

\noindent which depicts the balance of jet inertia with viscous forces.
\oo Using characteristic scales for jet momentum $\mathcal{M}_{\text{jet}} \sim \rho V_\text{jet} h_{\text{base}}^3$, velocity gradient $\partial_zv \sim V_{\text{jet}}/\delta_\eta$, and time $\tau_i \sim h_{\text{base}}/V_\text{jet}$, the force balance yields\bb

\begin{align}
	\rho V_{\text{jet}}^2 \sim \eta_{\text{effective}}\frac{V_{\text{jet}}}{\delta_\eta}.
	\label{scaling}
\end{align}

\noindent Here, $\delta_\eta$ represents the viscous length scale and the effective viscosity is

\begin{align}
	\label{eqn:etaEff}
  \eta_{\text{effective}} = 3\eta_s + 2G\lambda .
\end{align}

\noindent \oo Since polymers do not affect the flow before jet formation (\S~\ref{sec:polymers}), the jet Weber number remains constant at inception \citep{blanco2021jets},\bb

\begin{align}
	We_{\text{jet}} = \frac{\rho V_{\text{jet}}^2 \delta_\eta}{\gamma} =\,\text{constant}.
	\label{We}
\end{align}

\noindent Combining equations \eqref{scaling} and \eqref{We}, we get

\begin{align}
	V_{\text{jet}}  \sim \frac{\gamma}{\eta_{\text{effective}}}
	\label{velocityscale}
\end{align}

\noindent \oo analogous to Newtonian media but with modified viscosity \citep{gordillo2019capillary, blanco2020sea}.\bb

\begin{figure}
	\centering
	\includegraphics[width=\textwidth]{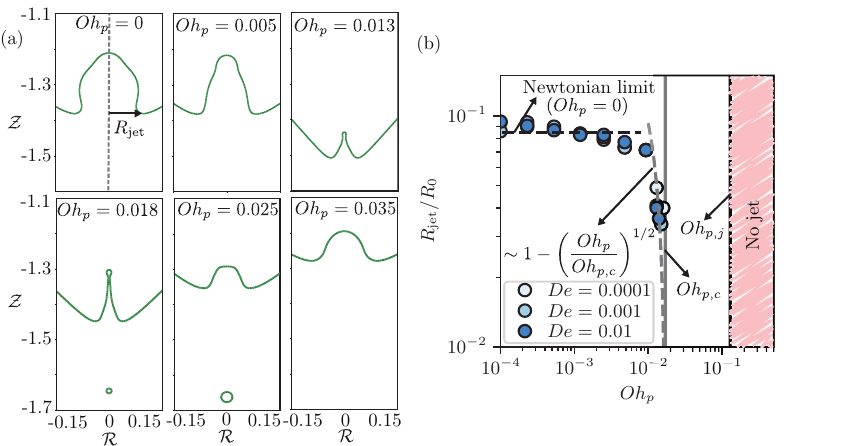}
	\caption{
		The capillary waves focus and collapse at the bottom of the cavity. (a) The inception of the jet after the collapse at different $Oh_p$ at $De = 0.001$ and $Oh_s = 0.01$. The radius of the jet at the base $R_{\text{jet}}$ decreases with $Oh_p$ until $Oh_{p,c} = 0.017$, beyond which bubbles are entrained and the jet radius increases.
		(b) Radius of jet $R_{\text{jet}}$ with $Oh_p$ at  $Oh_s = 0.01$ and different $De$. $R_{\text{jet}}$ remains close to the value at the Newtonian limit $Oh_p = 0$, and decreases sharply as it approaches $Oh_{p,c}$. Beyond $Oh_{p,j}$ jets are no longer observed.
	}
	\label{Rjet_Ohp}
\end{figure}

Figure~\ref{stressratio_De}(b) illustrates the jet velocity as a function of the effective Ohnesorge number

\begin{align}
   Oh_\text{effective} = 3Oh_s+2Oh_p
\end{align}

\noindent \DL{(reflecting equation~\eqref{eqn:etaEff})} at different $De$. We stress that the jet velocity varies in time \citep{deike2018dynamics,sanjay-2022-JFM,gordillo2023theory} and is maximum at its inception, which is the value that we report here.
For sufficiently large $Oh_{\text{effective}}$ and small $De$, we recover the scaling predicted in equation~\eqref{velocityscale}. However, as $De$ increases, the added elastic stresses cannot be directly substituted with Newtonian-like viscous stresses, and the underlying assumption fails, evident in the deviation of $V_{\text{jet}}$ from the prediction.

On the other hand, for small $Oh_{\text{effective}}$, $V_{\text{jet}}$ for all $De$ closely matches the corresponding speed in Newtonian liquids, as observed in figure~\ref{stressratio_De}b for $Oh_p = 0$. As $Oh_{p}$ increases, $V_{\text{jet}}$ also increases, reaching a maximum before decreasing and following equation~\eqref{velocityscale}.
Although \DL{the} capillary wave speed remains unaffected in the polymeric medium, increasing $Oh_p$ triggers elastic stresses in smaller wavelength capillary waves, which are promptly dissipated due to small $De$. Consequently, improved flow focusing occurs as the strongest undamped capillary wave survives, thus increasing $V_{\text{jet}}$. This behavior is analogous to the non-monotonicity observed and well-understood for Newtonian liquids at small $Oh_s$ \citep{duchemin2002jet, deike2018dynamics, gordillo2019capillary,sanjay-2022-JFM,yang2020multitude,gordillo2023theory}, further supporting the observation that \DL{polymeric liquid} exhibit a Newtonian-like viscous response in the zero $De$ limit.

To further quantify this behavior, figure~\ref{Rjet_Ohp}(a) illustrates jet features at inception for different $Oh_p$ at $De = 0.001$, while figure~\ref{Rjet_Ohp}(b) shows jet radius as a function of $Oh_p$ at different $De$.
\oo At small $Oh_p$, we observe that the jet radius maintains a value comparable to the Newtonian reference case (figure~\ref{Rjet_Ohp}a: $Oh_p = 0, 0.005$). This behavior is consistent with the $De \to 0$ limit, where polymeric additives primarily contribute enhanced effective viscosity. Since the jet radius determines the resulting drop size \citep{ganan2017revision,blanco2020sea}, this independence of jet radii in the low $Oh_p$ regime suggests minimal variation in droplet size distribution compared to Newtonian cases.\bb\,
As $Oh_p$ increases, we observe a pronounced reduction in jet width until reaching $Oh_{p,c} \approx 0.017$ (figure~\ref{Rjet_Ohp}a: $Oh_p = 0.013, 0.018$). At this critical value, the system transitions to a bubble entrainment regime (figure~\ref{Rjet_Ohp}a: $Oh_p = 0.018, 0.025$ \citet{gordillo2019capillary,blanco2020sea, rodriguez2023bubble}).
Interestingly, the prediction for Newtonian liquids applies well to viscoelastic liquids by substituting $Oh_p$ for $Oh_s$ (figure~\ref{Rjet_Ohp}b), particularly in the $De \to 0$ limit.
Beyond $Oh_{p,c}$, the jet radius becomes ill-defined as the jet gradually widens (figure~\ref{Rjet_Ohp}a: $Oh_p = 0.035$), first reaching the dropping transition at $Oh_{p,d} \approx 0.048$ (figures~\ref{fig:transitionDrops} and~\ref{Rjet_Ohp}b) and ultimately vanishing at $Oh_{p,j} \approx 0.129$ (figure~\ref{fig:transitionJets}b).

\section{Conclusion and outlook}
\label{sec:Conclusion}

This work elucidates the effects of viscoelasticity on Worthington jet formation and droplet ejection, \DL{by} contrasting \DL{it} with Newtonian fluid behavior. The process is governed by two key dimensionless parameters: the elastocapillary number $Ec$, comparing elastic and capillary forces, and the Deborah number $De$, relating the \DL{relaxation time of the polymeric liquid} to the inertiocapillary timescale. We identify three distinct regimes in viscoelastic media, analogous to Newtonian fluids: (i) jet formation with droplet ejection, (ii) jets without droplets, and (iii) complete jet suppression. However, the transitions between these regimes now depend on $Ec$ and $De$ rather than solely on the solvent Ohnesorge number $Oh_s$. Notably, while viscoelasticity significantly alters jet dynamics, it does not affect \DL{the} capillary wave speed.

Analysis across the $Ec$-$De$ phase space reveals markedly different behaviors in two limiting cases.
For polymeric liquids with permanent memory ($De \to \infty$), transitions occur at fixed $Ec$, independently of $De$. The jetting transition $Ec_j$ is independent of $Oh_s$, while the dropping transition $Ec_d$ exhibits strong $Oh_s$ dependence.
Remarkably, this infinite $De$ asymptote extends down to $De \approx 1$, where the polymer relaxation time becomes comparable to the process timescale.
\oo
Below this, for $De \sim \mathcal{O}(0.1)$, we observe a transition in scaling behavior, consistent with the Weissenberg number criterion $Wi \equiv De\sqrt{We_{\text{jet}}} \sim \mathcal{O}(1)$, where $We_{\text{jet}}$ is the jet Weber number (equation~\eqref{We}) that remains approximately constant due to negligible elastic effects during the initial shear flow \citep{blanco2021jets}.
\bb
Conversely, for polymeric liquids with poor memory ($De \to 0$), both transitions occur at constant polymeric Ohnesorge number $Oh_p = Ec \times De$, indicating that the addition of polymers introduces an excess viscous stress in this limit.
These transitions are independent of $Oh_s$.
Using a slender jet approach \citep{driessen2013stability,gordillo2020impulsive,eggers2015singularities}, we provide further insights into these transitions, examining the competition between elastic stresses and inertial flow focusing that governs jet formation and droplet ejection. This analysis helps to explain the observed scaling behaviors and transition criteria.

Our findings have important implications for understanding and controlling bubble bursting in viscoelastic fluids, with relevance to biological processes \citep{walls2017quantifying}, such as airborne disease transmission \citep{bourouiba2021fluid}, and industrial applications, such as inkjet printing \citep{lohse2022fundamental}.
\oo
The results highlight how polymer additives can dramatically alter spray formation, with intermediate values of $Ec$ and $De$ leading to smaller and faster droplets, whereas high values of $Ec$ and $De$ suppress droplet formation entirely \citep{kant2023bag}.
\bb
This work also opens several avenues for future research. Further investigation is needed into the universal behavior near the infinite $De$ asymptote, including the development of closed-form $De$ expansions for free surface flows \citep{sen2021retraction,francca2024elasto,sen2024elastocapillary,boyko2024flow,hinch2024fast}. The mechanism underlying the $Oh_s$ sensitivity of transition $Ec$ values at high $De$ requires further clarification. \VS{Additionally, extending our analysis to nonlinear viscoelastic models would provide valuable insights into the role of shear-thinning behavior and finite extensibility on bursting bubbles, addressing limitations of the current model \citep{zinelis2023transition,mckinley2002filament,snoeijer2020relationship}. This approach would allow quantification of discrepancies between experiments and simulations, often attributed to inherent issues with the Oldroyd-B model, thereby enhancing our understanding of viscoelastic jets \citep{gaillard2024does}.
Indeed, the numerical method developed here, freely available at \citet{Sanjay2024code}, provides a generalized framework readily adaptable to any model within the Oldroyd-B family of upper convective derivative models \citep{snoeijer2020relationship}.
Furthermore, as higher Bond numbers are observed in many scenarios \citep{walls2015jet, ghabache2014liquid, deike2018dynamics, krishnan2017scaling}, exploring their combined effect with viscoelasticity on the overall dynamics would provide valuable insights into such experiments \citep{rodriguez2023bubble}.
\oo
Indeed, a critical assumption of this work is the initial condition and its history, particularly for bubbles at liquid-gas interfaces in viscoelastic or elastoviscoplastic media. Our current work assumes the bubble has resided at the interface for a duration far exceeding the polymeric medium's relaxation time, ensuring elastic stresses have fully relaxed before bursting. This idealized scenario provides a well-defined starting point but may not fully capture experimental conditions \citep{cheny1996extravagant,deoclecio2023drop}.
\bb
Lastly, studying interactions of multiple bubbles \citep{singh2019numerical} at the liquid-gas free surface will provide further insights into pathogen transport.}

\VS{Extensions of this work could also explore coated bubbles \citep{dollet2019bubble, yang2023enhanced} or those with surface elasticity \citep{ji2023secondary}, and incorporate surfactants that alter bulk or interfacial properties \citep{constante2021dynamics,lohse2022fundamental,pierre2022influence,pico2024drop}.
Utilizing the current numerical framework to investigate the effects of bubble motion \citep{beris1985creeping,moschopoulos2021concept} and oscillations in viscoelastic media \citep{dollet2019bubble, oratis2024unifying} on overall dynamics before bursting would also be beneficial.
This model provides a general framework for studying both Newtonian viscous and non-Newtonian elastic effects. As a future perspective, it would be worthwhile to study phenomena such as wrinkling \citep{debregeas1998life, oratis2020new, davidovitch2024viscous} and buckling \citep{le2012buckling,timoshenko2012theory}, which occur in various viscoelastic systems \citep{schmalholz1999buckling,lee2024buckling,matoz2020wrinkle}.
By encompassing both viscous and elastic behaviors, this approach enables a comprehensive study of these interconnected instabilities, elucidating their underlying mechanisms and relationships as envisioned by \citet{stokes1845,rayleigh1896theory,taylor1969instability}. Moreover, integrating viscoelastic and elastoviscoplastic \citep{francca2024elasto,ari2024bursting} properties into recently developed analytical methods for capillary wave propagation and convergence, such as those by \citet{kayal2024focusing}, could yield a deeper theoretical understanding of the phenomenon.}

In conclusion, this study investigates and characterizes bubble bursting in viscoelastic media, interpreting the interplay between elastic, viscous, and capillary forces by moving in the $Oh_s$-$Ec$-$De$ phase space. 
\oo
As a starting point, we employed the Oldroyd-B constitutive model. While this choice elucidates the basic interplay of elasticity, viscosity, and capillarity, it does not capture shear-thinning effects or finite extensibility of polymer chains. Therefore, the predicted droplet sizes, jet thinning dynamics, and ultimate filament breakup must be interpreted with caution. More complex viscoelastic models (e.g., Giesekus, FENE-P) that incorporate finite extensibility and nonlinearities will likely alter certain details of our findings. Hence, our results should be viewed as a conceptual road map rather than definitive predictions. 
An essential extension of our study involves the experimental validation of the numerical results. Controlled laboratory studies using polymer solutions with known rheological properties are needed to assess the accuracy of the Oldroyd-B model in this parameter regime (also see appendix~\ref{app:accounting}). Such comparisons will help determine where the simplified assumptions fail and guide refinements, including the use of more realistic constitutive equations.

Despite these caveats, our study offers a foundation for understanding how viscoelasticity can either suppress or enhance droplet formation during bubble bursting. We hope this work will inspire future experiments and numerical explorations using more advanced rheological models, ultimately leading to a more complete and quantitative picture of viscoelastic bubble bursting across different application domains.
\bb\\

\noindent{\bf  Supplementary data\bf{.}} \label{SM} Supplementary material and movies are available at xxxx \\

\noindent{\bf  Code availability\bf{.}} The codes used in the present article are permanently available at \citet{Sanjay2024code}.\\

\noindent{\bf Acknowledgments\bf{.}} We would like to thank Andrea Prosperetti, Gareth McKinley, Jacco Snoeijer, Maziyar Jalaal, Uddalok Sen, and Vincent Bertin for discussions.\\

\noindent{\bf Funding\bf{.}} We acknowledge the funding from the MIST consortium. This publication is part of the project MIST with project number P20-35 of the research programme Perspectief, which is (partly) financed by the Dutch Research Council (NWO). We also acknowledge the NWO-Canon grant FIP-II grant. This work was carried out on the national e-infrastructure of SURFsara, a subsidiary of SURF cooperation, the collaborative ICT organization for Dutch education and research. This work was sponsored by NWO - Domain Science for the use of supercomputer facilities.\\

\noindent{\bf Declaration of Interests\bf{.}} The  authors report no conflict of interest. \\

\noindent{\bf Authors' ORCID\bf{.}} \\
A. K. Dixit \href{https://orcid.org/0000-0002-1544-6676}{orcid.org/0000-0002-1544-6676}\\
A. Oratis \href{https://orcid.org/0000-0002-6926-9302}{orcid.org/0000-0002-6926-9302} \\
K. Zinelis \href{https://orcid.org/0009-0009-4458-3221}{orcid.org/0009-0009-4458-3221} \\
D. Lohse \href{https://orcid.org/0000-0003-4138-2255}{orcid.org/0000-0003-4138-2255}\\
V. Sanjay \href{https://orcid.org/0000-0002-4293-6099}{orcid.org/0000-0002-4293-6099}\\

\appendix

\section{The Newtonian limit of bursting bubble dynamics}
\label{app:newtonian_limit}
\renewcommand{\thefigure}{\Alph{section}\,\arabic{figure}}
\setcounter{figure}{0}

\begin{figure}
	\includegraphics[width=\textwidth]{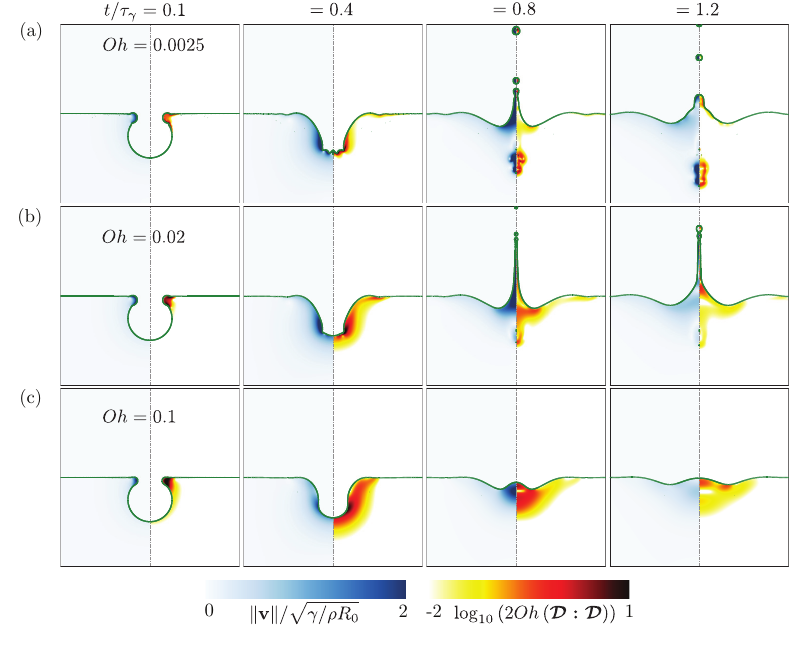}
	\caption{Temporal evolution of bubble cavity collapse in Newtonian liquid for $Oh_s =$ (a) $0.0025$, (b) $0.02$\AKD{,} and (c) $0.1$. The left panel represents the magnitude of the velocity field normalized by the inertiocapillary velocity, while the right panel shows the local viscous dissipation on a $\log_{10}$ scale. See also the supplementary movies SM3.}
	\label{facets_time_Oh_Newt}
\end{figure}

The dynamics of bursting bubbles in Newtonian media are solely dictated by the Ohnesorge number $Oh_s$ in the limit of very small bubbles (Bond number $Bo \ll 1$).
Figure~\ref{facets_time_Oh_Newt} illustrates representative cases at varying $Oh_s$ for $Bo = 0.001$. At low $Oh_s$, capillary waves propagate along the cavity, converging at its base to form a Worthington jet that subsequently fragments into droplets (figure~\ref{facets_time_Oh_Newt}a). In this limit, multiple undamped capillary waves collide at the cavity's bottom, generating a thick Worthington jet. Increasing $Oh_s$ dampens short-wavelength capillary waves, allowing the dominant wave to focus more effectively and produce a thinner jet. This explains the observed decrease in jet width with increasing $Oh_s$ \citep{gordillo2023theory}, until a critical $Oh_c \approx 0.03$ (at $Bo=0.001$) where the jet becomes extremely narrow, approaching a singularity \citep{blanco2020sea}. Concurrently, the size of the first ejected droplet diminishes with increasing $Oh_s$ \citep{gordillo2019capillary}.
As $Oh_s$ further increases, bubble entrainment occurs. Beyond $Oh_{s, d} = 0.0375$, vertical droplet ejection ceases; instead, the jet undergoes Rayleigh--Plateau instability, producing droplets that fall back into the pool \citep{blanco2020sea, walls2015jet, deike2018dynamics}. As $Oh_s$ increases ($Oh_s > 0.045$), viscous dissipation becomes more prominent, resulting in jet formation without droplet ejection (figure~\ref{facets_time_Oh_Newt}b). Further increase in $Oh_s$ beyond $Oh_{s,j} = 0.11$ completely suppresses jet formation (figure~\ref{facets_time_Oh_Newt}c, also see \citet{sanjay-2022-JFM}).

\oo
\section{A note on the range of control parameters considered in this work}
\label{app:accounting}
\renewcommand{\thetable}{\Alph{section}\,\arabic{table}}
\setcounter{table}{0}

In this appendix, we tabulate and compare the range of dimensionless parameters explored in this work with those available in the literature on viscoelastic effects in bubble bursting. Tables~\ref{tab:ExpOnlydim_numbers} and \ref{tab:dim_numbers} summarize the physical properties and corresponding dimensionless numbers from three representative experimental studies.

\begin{table}
	\begin{center}
		\begin{tabular}{lcccccc}
			&$c$ & $R$ & $\eta_s$  & $\lambda$ & $\eta_p$ & $G$ \\
			& (ppm) & ($\si{\milli\meter}$) & ($\si{\milli\pascal\second}$) & ($\si{\micro\second}$) & ($\si{\milli\pascal\second}$) & ($\si{\pascal}$) \\[3pt]
			\citet{cheny1996extravagant} & [0, 100]  & 7.5, 19  & 300   & N/A & [0, 18] & N/A \\
			\citet{rodriguez2023bubble} & [0, 350] & 1 & 1  & [0, 500] & [0, 0.5] & [0, 1] \\
			\citet{cabalganteeffect} & [0, 100] & 0.93 & 0.89 & [0, 700] & [0, 2] &[0, 1] \\
		\end{tabular}
		\caption{Representative values of physical parameters in polymer solution studies from three representative works on the Worthington jets from the literature. Across these studies, the density of the medium and its surface tension coefficient are roughly $1000\,\si{\kilogram}/\si{\cubic\meter}$ and $70\,\si{\milli\newton}/\si{\meter}$, respectively. N/A represents unavailable data. See table~\ref{tab:dim_numbers} for the estimates of dimensionless numbers using these properties.}
		\label{tab:ExpOnlydim_numbers}
	\end{center}
\end{table}

\begin{table}
	\begin{center}
		\def~{\hphantom{0}}
		\begin{tabular}{lccccc}
			& $Oh_s$ &  $De$ & $Ec$ & $Oh_p$ & $Bo$ \\[3pt]
			This work & [10$^{-3}$, $10^0$] & [0, $\infty$) & [0, 10$^3$] & [0, $\infty$) & $10^{-3}$ \\
			\citet{ari2024bursting} & [$10^{-3}, 10^{-2}$]  & [0, $10^2$] & [0, $10$] & [$10^{-3}, 10^{-2}$] & $10^{-3}$ \\
			\citet{cheny1996extravagant}  & $10^{-1}$ & N/A & N/A & [0, $10^{-2}$] & [$10, 10^2$] \\
			\citet{rodriguez2023bubble}  & $10^{-3}$  & [0, $10^{-1}$]& [0, $10^{-2}$] & [0, $10^{-3}$] & $10^{-1}$ \\
			\citet{cabalganteeffect} & $10^{-3}$ &  [0, $2 \times 10^{-1}$] & [0, $10^{-2}$] & [0, $10^{-2}$] & $10^{-1}$ \\
		\end{tabular}
		\caption{Representative values of dimensionless numbers in this work as compared to those from previous studies. For experimental studies, the dimensionless parameters are calculated using the properties in table~\ref{tab:ExpOnlydim_numbers}. For \citet{ari2024bursting}, we have only considered the limiting cases of zero yield-stress. We note that while experiments are naturally limited in their accessible parameter ranges, our numerical study explores a broader range to establish comprehensive scaling laws and regime transitions.}
		\label{tab:dim_numbers}
	\end{center}
\end{table}

Table~\ref{tab:ExpOnlydim_numbers} presents key physical parameters including polymer concentration ($c$), bubble radius ($R$), solvent viscosity ($\eta_s$), polymer relaxation time ($\lambda$), polymer contribution to viscosity ($\eta_p$), and elastic modulus ($G$). The corresponding dimensionless numbers are shown in Table~\ref{tab:dim_numbers}, where we compare our parameter space with both experimental and computational studies from the literature. Our work systematically explores a significantly broader range of these parameters compared to experimental studies, which are often constrained by practical limitations in achievable polymer concentrations and relaxation times. This comprehensive coverage allows us to identify universal scaling laws and regime transitions that may be challenging to observe experimentally.

The ranges explored in our numerical study suggest several promising directions for future experimental investigations. For instance, while moving in the $De$-$Ec$ parameter space, experiments could probe the robustness of our predicted transitions and scaling laws. Experimental studies would not only validate our computational findings but could also reveal additional physical mechanisms not captured by the Oldroyd-B model. We anticipate that trying new polymers and advances in characterization techniques \citep{gaillard2024beware} will continue to expand the experimentally accessible parameter space, enabling increasingly detailed comparisons between simulations and experiment.

\bb
\oo
\section{Grid sensitivity tests}
\label{app:gis}
\renewcommand{\thefigure}{\Alph{section}\,\arabic{figure}}
\setcounter{figure}{0}

This appendix assesses the grid independence of our numerical results by examining two important metrics: (i) the predicted droplet size and (ii) the regime transitions. Ensuring grid convergence is crucial, especially if interface ruptures due to finite grid resolution in our numerical code \citep{lohse-2020-pnas,chirco2022manifold,kant2023bag}. 

\begin{figure}
	\centering
	\includegraphics[width=\textwidth]{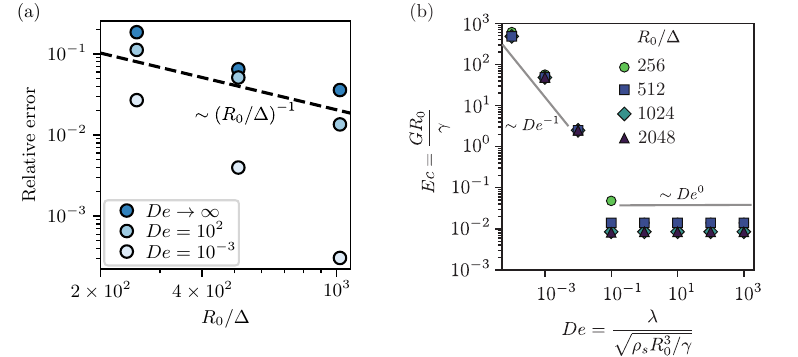}
	\caption{ \rev{(a) The relative error in predicted droplet size versus the number of grid points per bubble radius, $R_0/\Delta$, at $De \to \infty$, $De = 10^2$, and $De = 10^{-3}$. The dashed line indicates a scaling of $(R_0/\Delta)^{-1}$, demonstrating approximately first-order convergence for large $De$ cases. The relative error for small $De$ is lower as the elastic stresses are less prominent compared to large $De$.} (b) Dependence of the critical elastocapillary number $Ec_d$ at the dropping transition on the Deborah number $De$ for different grid resolutions ($R_0/\Delta = 256, 512, 1024, 2048$). The scaling behaviors $Ec_d \sim De^{-1}$ as $De \to 0$ and $Ec_d \sim De^0$ as $De \to \infty$ remain unchanged beyond $R_0/\Delta = 1024$.}
	\label{fig:gis}
\end{figure}

Figure~\ref{fig:gis}(a) shows the relative error in predicted droplet size as a function of the number of grid points per initial bubble radius $R_0/\Delta$, where $\Delta$ is the minimum grid size. We focus on $De \to \infty$ as this case is particularly demanding, featuring slender filaments due to viscoelastic stresses. The error is calculated relative to the finest resolution ($R_0/\Delta = 2048$). 
The data exhibit approximately first-order convergence, indicated by the dashed line scaling as $(R_0/\Delta)^{-1}$. 
For our standard resolution of $R_0/\Delta = 512$, the relative error is approximately 6\%, decreasing to about 3\% at $R_0/\Delta = 1024$.

While droplet size convergence demonstrates improved numerical accuracy with increasing resolution, the determination of regime transitions between different flow behaviors provides an even more stringent test. These transitions are highly sensitive to the details of jet breakup. Figure~\ref{fig:gis}(b) displays the dimensionless elastocapillary number $Ec$ at the transition boundary for different grid resolutions. We find that for $(R_0/\Delta) \geq 1024$, the transition curves do not change, confirming that the scaling behaviors previously identified -- namely $Ec_d \sim De^{-1}$ for $De \ll 1$ and $Ec_d \sim De^0$ for $De \gg 1$ -- are robustly reproduced across all grid resolutions tested.

\section{Deviation from the Newtonian asymptote}\label{app:devNewt}
\renewcommand{\thefigure}{\Alph{section}\,\arabic{figure}}
\setcounter{figure}{0}
\begin{figure}
    \centering
    \includegraphics[width=\textwidth]{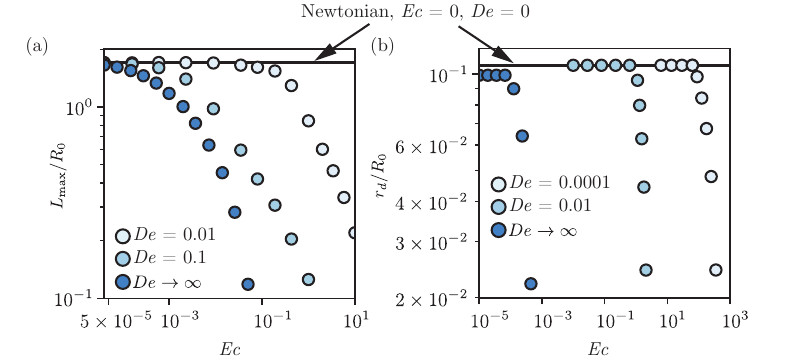}
    \caption{{\oo Comparison of (a) maximum jet length $L_{\text{max}}/R_0$ against $Ec$ at various $De$ at fixed representative cases of $Oh_s=0.05$ and (b) first droplet size $r_d/R_0$  against $Ec$ at various $De$ fixed at $Oh_s=0.001$. The horizontal lines indicate the Newtonian reference values (obtained at $Ec=0$). At small $Ec$, both $L_{\text{max}}$ and $r_d$ coincide with their Newtonian counterparts, demonstrating negligible viscoelastic influence. As $Ec$ increases beyond critical values, significant deviations from the Newtonian limits emerge, with the degree of departure depending on $De$. These results quantify the onset and magnitude of elastic effects relative to the Newtonian baseline, providing a clear framework for interpreting viscoelastic modifications to bursting bubble dynamics.\bb}}
    \label{fig:devNewt}
\end{figure}

In the main text, we showed that for small elastocapillary numbers $Ec$, the droplet size $r_d$ and jet length $L_{\text{max}}$ closely match those of the Newtonian case at arbitary $Oh_s$ (solvent Ohnesorge number) and $De$ (Deborah number). Only when $Ec$ approaches or exceeds critical values do we observe significant departures from the Newtonian reference.

Figure~\ref{fig:devNewt} quantifies these deviations by comparing both the maximum jet length $L_{\text{max}}$ (Figure~\ref{fig:devNewt}a) and the first droplet size $r_d$ (Figure~\ref{fig:devNewt}b) against $Ec$ at various $De$, in the limit of $Oh_s \ll 1$. The symbols represent numerical results for the viscoelastic system, while the horizontal lines mark the corresponding Newtonian asymptotes (i.e., $r_d$ and $L_{\text{max}}$ values obtained at $Ec=0$). For small $Ec$, both $r_d$ and $L_{\text{max}}$ are invariant, indicating that viscoelastic stresses are negligible in this range. As $Ec$ increases and approaches the critical thresholds identified in \S~\ref{sec:regimes}, deviations emerge, ultimately leading to suppressed jetting or droplet formation.

Notably, the critical $Ec$ value at which $r_d$ and $L_{\text{max}}$ deviate from their Newtonian counterparts depends on $De$. For high $De$, even a moderate increase in $Ec$ can trigger significant changes, reflecting the persistent elastic memory in the fluid. In contrast, for $De \ll 1$, where the polymeric stresses relax rapidly, larger $Ec$ values are necessary to produce noticeable departures from Newtonian behavior. Similarly, the Newtonian limit is readily recovered by reducing either $Ec$ or $De$ to zero.

These results highlight that any interpretation of viscoelastic bubble-bursting dynamics should be framed with reference to the Newtonian baselines (either $De = 0$ or $Ec = 0$). By systematically mapping out these deviations, one can pinpoint the onset of non-Newtonian behavior and interpret observed jetting or droplet formation regimes as outcomes of either weak or strong elastic effects, all benchmarked against the Newtonian scenario.

\bb

\bibliographystyle{jfm}
%% Note the spaces between the initials
\bibliography{bubbleBurstingVE}

\begin{thebibliography}{158}
\expandafter\ifx\csname natexlab\endcsname\relax\def\natexlab#1{#1}\fi
\def\au#1{#1} \def\ed#1{#1} \def\yr#1{#1}\def\at#1{#1}\def\jt#1{\textit{#1}}
  \def\bt#1{#1}\def\bvol#1{\textbf{#1}} \def\vol#1{#1} \def\pg#1{#1}
  \def\publ#1{#1}\def\arxiv#1{#1}\def\org#1{#1}\def\st#1{\textit{#1}}

\bibitem[Afkhami {\em et~al.\/}(2018)Afkhami, Buongiorno, Guion, Popinet,
  Saade, Scardovelli \& Zaleski]{afkhamiTransitionNumericalModel2018}
{\sc \au{Afkhami, S.}, \au{Buongiorno, J.}, \au{Guion, A.}, \au{Popinet, S.},
  \au{Saade, Y.}, \au{Scardovelli, R.} \& \au{Zaleski, S.}} \yr{2018}
  \at{Transition in a numerical model of contact line dynamics and forced
  dewetting}.  \jt{J. Comput. Phys.}  \bvol{374},  \pg{1061--1093}.

\bibitem[Alves {\em et~al.\/}(2021)Alves, Oliveira \&
  Pinho]{alves2021numerical}
{\sc \au{Alves, M.~A.}, \au{Oliveira, P.~J.} \& \au{Pinho, F.~T.}} \yr{2021}
  \at{Numerical methods for viscoelastic fluid flows}.  \jt{Annu. Rev. Fluid
  Mech.}  \bvol{53}~(1),  \pg{509--541}.

\bibitem[Anna \& McKinley(2001)]{anna2001elasto}
{\sc \au{Anna, S.~L.} \& \au{McKinley, G.~H.}} \yr{2001}  \at{Elasto-capillary
  thinning and breakup of model elastic liquids}.  \jt{J. Rheol.}
  \bvol{45}~(1),  \pg{115--138}.

\bibitem[Balasubramanian {\em et~al.\/}(2024)Balasubramanian, Sanjay, Jalaal,
  Vinuesa \& Tammisola]{ari2024bursting}
{\sc \au{Balasubramanian, A.~G.}, \au{Sanjay, V.}, \au{Jalaal, M.},
  \au{Vinuesa, R.} \& \au{Tammisola, O.}} \yr{2024}  \at{Bursting bubble in an
  elastoviscoplastic medium}.  \jt{Journal of Fluid Mechanics}  \bvol{1001},
  \pg{A9}.

\bibitem[Balci {\em et~al.\/}(2011)Balci, Thomases, Renardy \&
  Doering]{balci2011symmetric}
{\sc \au{Balci, N.}, \au{Thomases, B.}, \au{Renardy, M.} \& \au{Doering,
  C.~R.}} \yr{2011}  \at{Symmetric factorization of the conformation tensor in
  viscoelastic fluid models}.  \jt{J. Non-Newtonian Fluid Mech.}
  \bvol{166}~(11),  \pg{546--553}.

\bibitem[Bartlett {\em et~al.\/}(2023)Bartlett, Oratis, Santin \&
  Bird]{bartlett2023universal}
{\sc \au{Bartlett, C.}, \au{Oratis, A.~T.}, \au{Santin, M.} \& \au{Bird,
  J.~C.}} \yr{2023}  \at{Universal non-monotonic drainage in large bare viscous
  bubbles}.  \jt{Nat. Commun.}  \bvol{14}~(1),  \pg{877}.

\bibitem[Bergmann {\em et~al.\/}(2009)Bergmann, {van der Meer}, Gekle, {van der
  Bos} \& Lohse]{bergmann2009controlled}
{\sc \au{Bergmann, R.}, \au{{van der Meer}, D.}, \au{Gekle, S.}, \au{{van der
  Bos}, A.} \& \au{Lohse, D.}} \yr{2009}  \at{Controlled impact of a disk on a
  water surface: cavity dynamics}.  \jt{J. Fluid Mech.}  \bvol{633},
  \pg{381--409}.

\bibitem[Bergmann {\em et~al.\/}(2006)Bergmann, {van der Meer}, Stijnman,
  Sandtke, Prosperetti \& Lohse]{bergmann2006giant}
{\sc \au{Bergmann, R.}, \au{{van der Meer}, D.}, \au{Stijnman, M.},
  \au{Sandtke, M.}, \au{Prosperetti, A.} \& \au{Lohse, D.}} \yr{2006}
  \at{Giant bubble pinch-off}.  \jt{Phys. Rev. Lett.}  \bvol{96}~(15),
  \pg{154505}.

\bibitem[Beris {\em et~al.\/}(1985)Beris, Tsamopoulos, Armstrong \&
  Brown]{beris1985creeping}
{\sc \au{Beris, A.~N.}, \au{Tsamopoulos, J.~A.}, \au{Armstrong, R.~C.} \&
  \au{Brown, R.~A.}} \yr{1985}  \at{Creeping motion of a sphere through a
  bingham plastic}.  \jt{J. Fluid Mech.}  \bvol{158},  \pg{219--244}.

\bibitem[Berny {\em et~al.\/}(2020)Berny, Deike, S{\'e}on \&
  Popinet]{berny2020role}
{\sc \au{Berny, A.}, \au{Deike, L.}, \au{S{\'e}on, T.} \& \au{Popinet, S.}}
  \yr{2020}  \at{Role of all jet drops in mass transfer from bursting bubbles}.
   \jt{Phys. Rev. Fluids}  \bvol{5}~(3),  \pg{033605}.

\bibitem[Berny {\em et~al.\/}(2021)Berny, Popinet, S{\'e}on \&
  Deike]{berny2021statistics}
{\sc \au{Berny, A.}, \au{Popinet, S.}, \au{S{\'e}on, T.} \& \au{Deike, L.}}
  \yr{2021}  \at{Statistics of jet drop production}.  \jt{Geophys. Res. Lett.}
  \bvol{48}~(10),  \pg{e2021GL092919}.

\bibitem[Bertin {\em et~al.\/}(2024)Bertin, Sanjay, Oratis \&
  Snoeijer]{BertinSanjay2024}
{\sc \au{Bertin, V.}, \au{Sanjay, V.}, \au{Oratis, A.~T.} \& \au{Snoeijer,
  J.~H.}} \yr{2024}  \at{Elastic {Taylor--Culick} retractions}.  \jt{Working
  paper} .

\bibitem[Bird {\em et~al.\/}(1980)Bird, Dotson \& Johnson]{bird1980polymer}
{\sc \au{Bird, R.~B.}, \au{Dotson, P.~J.} \& \au{Johnson, N.~L.}} \yr{1980}
  \at{Polymer solution rheology based on a finitely extensible bead---spring
  chain model}.  \jt{J. Non-Newton. Fluid Mech.}  \bvol{7}~(2-3),
  \pg{213--235}.

\bibitem[Bird {\em et~al.\/}(1977)Bird, Armstrong \&
  Hassager]{bird1977dynamics}
{\sc \au{Bird, R.~R.}, \au{Armstrong, R.~C.} \& \au{Hassager, O.}} \yr{1977}
  {\em Dynamics of Polymeric Liquids, Volume 1: Fluid Mechanics\/}.
  \publ{Wiley}.

\bibitem[Blanco-Rodr{\'\i}guez \& Gordillo(2020)]{blanco2020sea}
{\sc \au{Blanco-Rodr{\'\i}guez, F.~J.} \& \au{Gordillo, J.~M.}} \yr{2020}
  \at{On the sea spray aerosol originated from bubble bursting jets}.  \jt{J.
  Fluid Mech.}  \bvol{886},  \pg{R2}.

\bibitem[Blanco-Rodr{\'\i}guez \& Gordillo(2021)]{blanco2021jets}
{\sc \au{Blanco-Rodr{\'\i}guez, F.~J.} \& \au{Gordillo, J.~M.}} \yr{2021}
  \at{On the jets produced by drops impacting a deep liquid pool and by
  bursting bubbles}.  \jt{J. Fluid Mech.}  \bvol{916},  \pg{A37}.

\bibitem[Bogy(1979)]{bogy1979drop}
{\sc \au{Bogy, D.~B.}} \yr{1979}  \at{Drop formation in a circular liquid jet}.
   \jt{Annu. Rev. Fluid Mech.}  \bvol{11},  \pg{207--228}.

\bibitem[Bouillant {\em et~al.\/}(2022)Bouillant, Dekker, Hack \&
  Snoeijer]{bouillant2022rapid}
{\sc \au{Bouillant, A.}, \au{Dekker, P.~J.}, \au{Hack, M.~A.} \& \au{Snoeijer,
  J.~H.}} \yr{2022}  \at{Rapid viscoelastic spreading}.  \jt{Phys. Rev. Fluids}
   \bvol{7}~(12),  \pg{123604}.

\bibitem[Boulton-Stone \& Blake(1993)]{boulton1993gas}
{\sc \au{Boulton-Stone, J.~M.} \& \au{Blake, J.~R.}} \yr{1993}  \at{Gas bubbles
  bursting at a free surface}.  \jt{J. Fluid Mech.}  \bvol{254},
  \pg{437--466}.

\bibitem[Bourouiba(2021)]{bourouiba2021fluid}
{\sc \au{Bourouiba, L.}} \yr{2021}  \at{The fluid dynamics of disease
  transmission}.  \jt{Annu. Rev. Fluid Mech.}  \bvol{53},  \pg{473--508}.

\bibitem[Bousfield {\em et~al.\/}(1986)Bousfield, Keunings, Marrucci \&
  Denn]{bousfield1986nonlinear}
{\sc \au{Bousfield, D.~W.}, \au{Keunings, R.}, \au{Marrucci, G.} \& \au{Denn,
  M.~M.}} \yr{1986}  \at{Nonlinear analysis of the surface tension driven
  breakup of viscoelastic filaments}.  \jt{J. Non-Newtonian Fluid Mech.}
  \bvol{21}~(1),  \pg{79--97}.

\bibitem[Boyko {\em et~al.\/}(2024)Boyko, Hinch \& Stone]{boyko2024flow}
{\sc \au{Boyko, E.}, \au{Hinch, J.} \& \au{Stone, H.~A.}} \yr{2024}  \at{Flow
  of an oldroyd-b fluid in a slowly varying contraction: theoretical results
  for arbitrary values of deborah number in the ultra-dilute limit}.  \jt{J.
  Fluid Mech.}  \bvol{988},  \pg{A10}.

\bibitem[Boyko \& Stone(2024)]{boyko2024perspective}
{\sc \au{Boyko, E.} \& \au{Stone, H.~A.}} \yr{2024}  \at{Perspective on the
  description of viscoelastic flows via continuum elastic dumbbell models}.
  \jt{J. Eng. Math.}  \bvol{147}~(1),  \pg{1--18}.

\bibitem[Brackbill {\em et~al.\/}(1992)Brackbill, Kothe \&
  Zemach]{brackbill1992continuum}
{\sc \au{Brackbill, J.~U.}, \au{Kothe, D.~B.} \& \au{Zemach, C.}} \yr{1992}
  \at{A continuum method for modeling surface tension}.  \jt{J. Comput. Phys.}
  \bvol{100}~(2),  \pg{335--354}.

\bibitem[Cabalgante-Corrales {\em et~al.\/}(2025)Cabalgante-Corrales,
  Mu{\~n}oz-S{\'a}nchez, L{\'o}pez-Herrera, Cabezas, Vega \&
  Montanero]{cabalganteeffect}
{\sc \au{Cabalgante-Corrales, E.}, \au{Mu{\~n}oz-S{\'a}nchez, B.~N.},
  \au{L{\'o}pez-Herrera, J.~M.}, \au{Cabezas, M.~G.}, \au{Vega, E.~J.} \&
  \au{Montanero, J.~M.}} \yr{2025}  \at{Effect of the polymer viscosity and
  relaxation time on the worthington jet produced by bubble bursting in weakly
  viscoelastic liquids}.  \jt{Int. J. Multiph. Flow}  \bvol{184},  \pg{105095}.

\bibitem[Chang {\em et~al.\/}(1999)Chang, Demekhin \&
  Kalaidin]{chang1999iterated}
{\sc \au{Chang, H.}, \au{Demekhin, E.~A.} \& \au{Kalaidin, E.}} \yr{1999}
  \at{Iterated stretching of viscoelastic jets}.  \jt{Phys. Fluids}
  \bvol{11}~(7),  \pg{1717--1737}.

\bibitem[Chen(1991)]{chen1991interfacial}
{\sc \au{Chen, K.}} \yr{1991}  \at{Interfacial instability due to elastic
  stratification in concentric coextrusion of two viscoelastic fluids}.  \jt{J.
  Non-Newtonian Fluid Mech.}  \bvol{40}~(2),  \pg{155--175}.

\bibitem[Cheny \& Walters(1996)]{cheny1996extravagant}
{\sc \au{Cheny, J.~M.} \& \au{Walters, K.}} \yr{1996}  \at{Extravagant
  viscoelastic effects in the worthington jet experiment}.  \jt{J. Non-Newton.
  Fluid Mech.}  \bvol{67},  \pg{125--135}.

\bibitem[Chirco {\em et~al.\/}(2022)Chirco, Maarek, Popinet \&
  Zaleski]{chirco2022manifold}
{\sc \au{Chirco, L.}, \au{Maarek, J.}, \au{Popinet, S.} \& \au{Zaleski, S.}}
  \yr{2022}  \at{Manifold death: a volume of fluid implementation of controlled
  topological changes in thin sheets by the signature method}.  \jt{J. Comput.
  Phys.}  \bvol{467},  \pg{111468}.

\bibitem[Clasen {\em et~al.\/}(2006)Clasen, Eggers, Fontelos, Li \&
  McKinley]{clasen2006beads}
{\sc \au{Clasen, C.}, \au{Eggers, J.}, \au{Fontelos, M.~A.}, \au{Li, J.} \&
  \au{McKinley, G.~H.}} \yr{2006}  \at{The beads-on-string structure of
  viscoelastic threads}.  \jt{J. Fluid Mech.}  \bvol{556},  \pg{283--308}.

\bibitem[Constante-Amores {\em et~al.\/}(2021)Constante-Amores, Kahouadji,
  Batchvarov, Shin, Chergui, Juric \& Matar]{constante2021dynamics}
{\sc \au{Constante-Amores, C.~R.}, \au{Kahouadji, L.}, \au{Batchvarov, A.},
  \au{Shin, S.}, \au{Chergui, J.}, \au{Juric, D.} \& \au{Matar, O.~K.}}
  \yr{2021}  \at{Dynamics of a surfactant-laden bubble bursting through an
  interface}.  \jt{J. Fluid Mech.}  \bvol{911},  \pg{A57}.

\bibitem[Culick(1960)]{culick-1960-japplphys}
{\sc \au{Culick, F. E.~C.}} \yr{1960}  \at{Comments on a ruptured soap film}.
  \jt{J. Appl. Phys.}  \bvol{31},  \pg{1128--1129}.

\bibitem[Daneshi \& Frigaard(2024)]{daneshi2024growth}
{\sc \au{Daneshi, M.} \& \au{Frigaard, I.~A.}} \yr{2024}  \at{Growth and static
  stability of bubble clouds in yield stress fluids}.  \jt{J. Non-Newton. Fluid
  Mech.}  \bvol{327},  \pg{105217}.

\bibitem[Dasouqi {\em et~al.\/}(2022)Dasouqi, Ghossein \&
  Murphy]{dasouqi2022effect}
{\sc \au{Dasouqi, A.~A.}, \au{Ghossein, J.} \& \au{Murphy, D.~W.}} \yr{2022}
  \at{The effect of liquid properties on the release of gas from bursting
  bubbles}.  \jt{Exp. Fluids}  \bvol{63}~(1),  \pg{39}.

\bibitem[Davidovitch \& Klein(2024)]{davidovitch2024viscous}
{\sc \au{Davidovitch, B.} \& \au{Klein, A.}} \yr{2024}  \at{How viscous bubbles
  collapse: Topological and symmetry-breaking instabilities in curvature-driven
  hydrodynamics}.  \jt{{Proc. Natl. Acad. Sci. USA}}  \bvol{121}~(32),
  \pg{e2310195121}.

\bibitem[Davoodi {\em et~al.\/}(2018)Davoodi, Lerouge, Norouzi \&
  Poole]{davoodi2018secondary}
{\sc \au{Davoodi, M.}, \au{Lerouge, S.}, \au{Norouzi, M.} \& \au{Poole, R.~J.}}
  \yr{2018}  \at{Secondary flows due to finite aspect ratio in inertialess
  viscoelastic taylor--couette flow}.  \jt{J. Fluid Mech.}  \bvol{857},
  \pg{823--850}.

\bibitem[{de Gennes}(1974)]{de1974coil}
{\sc \au{{de Gennes}, P.-G.}} \yr{1974}  \at{Coil-stretch transition of dilute
  flexible polymers under ultrahigh velocity gradients}.  \jt{J. Chem. Phys.}
  \bvol{60}~(12),  \pg{5030--5042}.

\bibitem[{de Leeuw} {\em et~al.\/}(2011){de Leeuw}, Andreas, Anguelova,
  Fairall, Lewis, O'Dowd, Schulz \& Schwartz]{de2011production}
{\sc \au{{de Leeuw}, G.}, \au{Andreas, E.~L.}, \au{Anguelova, M.~D.},
  \au{Fairall, C.~W.}, \au{Lewis, E.~R.}, \au{O'Dowd, C.}, \au{Schulz, M.} \&
  \au{Schwartz, S.~E.}} \yr{2011}  \at{Production flux of sea spray aerosol}.
  \jt{Rev. Geophys.}  \bvol{49}~(2).

\bibitem[Debr{\'e}geas {\em et~al.\/}(1998)Debr{\'e}geas, {de Gennes} \&
  {Brochard-Wyart}]{debregeas1998life}
{\sc \au{Debr{\'e}geas, G.~D.}, \au{{de Gennes}, P.-G.} \&
  \au{{Brochard-Wyart}, F.}} \yr{1998}  \at{The life and death of ``bare"
  viscous bubbles}.  \jt{Science}  \bvol{279}~(5357),  \pg{1704--1707}.

\bibitem[Deike(2022)]{deike2022mass}
{\sc \au{Deike, L.}} \yr{2022}  \at{Mass transfer at the ocean--atmosphere
  interface: the role of wave breaking, droplets, and bubbles}.  \jt{Annu. Rev.
  Fluid Mech.}  \bvol{54},  \pg{191--224}.

\bibitem[Deike {\em et~al.\/}(2018)Deike, Ghabache, Liger-Belair, Das, Zaleski,
  Popinet \& S{\'e}on]{deike2018dynamics}
{\sc \au{Deike, L.}, \au{Ghabache, E.}, \au{Liger-Belair, G.}, \au{Das, A.~K.},
  \au{Zaleski, S.}, \au{Popinet, S.} \& \au{S{\'e}on, T.}} \yr{2018}
  \at{Dynamics of jets produced by bursting bubbles}.  \jt{Phys. Rev. Fluids}
  \bvol{3}~(1),  \pg{013603}.

\bibitem[Dekker {\em et~al.\/}(2022)Dekker, Hack, Tewes, Datt, Bouillant \&
  Snoeijer]{dekker2022elasticity}
{\sc \au{Dekker, P.~J.}, \au{Hack, M.~A.}, \au{Tewes, W.}, \au{Datt, C.},
  \au{Bouillant, A.} \& \au{Snoeijer, J.~H.}} \yr{2022}  \at{When elasticity
  affects drop coalescence}.  \jt{Phys. Rev. Lett.}  \bvol{128}~(2),
  \pg{028004}.

\bibitem[Deoclecio {\em et~al.\/}(2023)Deoclecio, Soares \&
  Popinet]{deoclecio2023drop}
{\sc \au{Deoclecio, L. H.~P.}, \au{Soares, E.~J.} \& \au{Popinet, S.}}
  \yr{2023}  \at{Drop rise and interfacial coalescence initiation in bingham
  materials}.  \jt{J. Non-Newton. Fluid}  \bvol{319},  \pg{105075}.

\bibitem[Dollet {\em et~al.\/}(2019)Dollet, Marmottant \&
  Garbin]{dollet2019bubble}
{\sc \au{Dollet, B.}, \au{Marmottant, P.} \& \au{Garbin, V.}} \yr{2019}
  \at{Bubble dynamics in soft and biological matter}.  \jt{Annu. Rev. Fluid
  Mech.}  \bvol{51}~(1),  \pg{331--355}.

\bibitem[Driessen {\em et~al.\/}(2013)Driessen, Jeurissen, Wijshoff, Toschi \&
  Lohse]{driessen2013stability}
{\sc \au{Driessen, T.~J.}, \au{Jeurissen, R.}, \au{Wijshoff, H.}, \au{Toschi,
  F.} \& \au{Lohse, D.}} \yr{2013}  \at{Stability of viscous long liquid
  filaments}.  \jt{Phys. Fluids}  \bvol{25}~(6).

\bibitem[Dubitsky {\em et~al.\/}(2023{\natexlab{{\em a\/}}})Dubitsky, McRae \&
  Bird]{dubitsky2023enrichment}
{\sc \au{Dubitsky, L.}, \au{McRae, O.} \& \au{Bird, J.~C.}}
  \yr{2023{\natexlab{{\em a\/}}}}  \at{Enrichment of scavenged particles in jet
  drops determined by bubble size and particle position}.  \jt{Phys. Rev.
  Lett.}  \bvol{130}~(5),  \pg{054001}.

\bibitem[Dubitsky {\em et~al.\/}(2023{\natexlab{{\em b\/}}})Dubitsky, Stokes,
  Deane \& Bird]{dubitsky2023effects}
{\sc \au{Dubitsky, L.}, \au{Stokes, M.~D.}, \au{Deane, G.~B.} \& \au{Bird,
  J.~C.}} \yr{2023{\natexlab{{\em b\/}}}}  \at{Effects of salinity beyond
  coalescence on submicron aerosol distributions}.  \jt{J. Geophys. Res.
  Atmos.}  \bvol{128}~(10),  \pg{e2022JD038222}.

\bibitem[Duchemin {\em et~al.\/}(2002)Duchemin, Popinet, Josserand \&
  Zaleski]{duchemin2002jet}
{\sc \au{Duchemin, L.}, \au{Popinet, S.}, \au{Josserand, C.} \& \au{Zaleski,
  S.}} \yr{2002}  \at{Jet formation in bubbles bursting at a free surface}.
  \jt{Phys. Fluids}  \bvol{14}~(9),  \pg{3000--3008}.

\bibitem[Eggers(1997)]{eggers1997nonlinear}
{\sc \au{Eggers, J.}} \yr{1997}  \at{Nonlinear dynamics and breakup of
  free-surface flows}.  \jt{Rev. Mod. Phys.}  \bvol{69}~(3),  \pg{865}.

\bibitem[Eggers \& Fontelos(2015)]{eggers2015singularities}
{\sc \au{Eggers, J.} \& \au{Fontelos, M.~A.}} \yr{2015} {\em Singularities:
  formation, structure, and propagation\/}, ,  \vol{vol.~53}.  \publ{Cambridge
  University Press}.

\bibitem[Eggers {\em et~al.\/}(2020)Eggers, Herrada \&
  Snoeijer]{eggers2020self}
{\sc \au{Eggers, J.~A.}, \au{Herrada, M.~A.} \& \au{Snoeijer, J.~H.}} \yr{2020}
   \at{Self-similar breakup of polymeric threads as described by the oldroyd-b
  model}.  \jt{J. Fluid Mech.}  \bvol{887},  \pg{A19}.

\bibitem[Eggers {\em et~al.\/}(2025)Eggers, Sprittles \&
  Snoeijer]{snoeijer2025coalescence}
{\sc \au{Eggers, J.~H.}, \au{Sprittles, J.~E.} \& \au{Snoeijer, J.}} \yr{2025}
  \at{Coalescence dynamics}.  \jt{Annu. Rev. Fluid Mech.} .

\bibitem[Fattal \& Kupferman(2004)]{fattal2004constitutive}
{\sc \au{Fattal, R.} \& \au{Kupferman, R.}} \yr{2004}  \at{Constitutive laws
  for the matrix-logarithm of the conformation tensor}.  \jt{J. Non-Newtonian
  Fluid Mech.}  \bvol{123}~(2-3),  \pg{281--285}.

\bibitem[Fraggedakis {\em et~al.\/}(2016)Fraggedakis, Pavlidis, Dimakopoulos \&
  Tsamopoulos]{fraggedakis2016velocity}
{\sc \au{Fraggedakis, D.}, \au{Pavlidis, M.}, \au{Dimakopoulos, Y.} \&
  \au{Tsamopoulos, J.}} \yr{2016}  \at{On the velocity discontinuity at a
  critical volume of a bubble rising in a viscoelastic fluid}.  \jt{J. Fluid
  Mech.}  \bvol{789},  \pg{310--346}.

\bibitem[Fran{\c{c}}a {\em et~al.\/}(2024)Fran{\c{c}}a, Jalaal \&
  Oishi]{francca2024elasto}
{\sc \au{Fran{\c{c}}a, H.~L.}, \au{Jalaal, M.} \& \au{Oishi, C.~M.}} \yr{2024}
  \at{Elasto-viscoplastic spreading: From plastocapillarity to
  elastocapillarity}.  \jt{Phys. Rev. Res.}  \bvol{6}~(1),  \pg{013226}.

\bibitem[Fullana {\em et~al.\/}(2024)Fullana, Kulkarni, Fricke, Popinet,
  Afkhami, Bothe \& Zaleski]{fullanaConsistentTreatmentDynamic2024}
{\sc \au{Fullana, T.}, \au{Kulkarni, Y.}, \au{Fricke, M.athis}, \au{Popinet,
  S.}, \au{Afkhami, S.}, \au{Bothe, D.} \& \au{Zaleski, S}} \yr{2024} A
  consistent treatment of dynamic contact angles in the sharp-interface
  framework with the generalized {{Navier}} boundary condition,  \arxiv{arXiv:
  2411.10762}.

\bibitem[Gaillard {\em et~al.\/}(2024{\natexlab{{\em a\/}}})Gaillard, Herrada,
  Deblais, Eggers \& Bonn]{gaillard2024beware}
{\sc \au{Gaillard, A.}, \au{Herrada, M.~A.}, \au{Deblais, A.}, \au{Eggers, J.}
  \& \au{Bonn, D.}} \yr{2024{\natexlab{{\em a\/}}}}  \at{Beware of {CaBER}:
  Filament thinning rheometry does not always give `the'relaxation time of
  polymer solutions}.  \jt{Phys. Rev. Fluids}  \bvol{9}~(7),  \pg{073302}.

\bibitem[Gaillard {\em et~al.\/}(2024{\natexlab{{\em b\/}}})Gaillard, Herrada,
  Deblais, {van Poelgeest}, Laruelle, Eggers \& Bonn]{gaillard2024does}
{\sc \au{Gaillard, A.}, \au{Herrada, M.~A.}, \au{Deblais, A.}, \au{{van
  Poelgeest}, C.}, \au{Laruelle, L.}, \au{Eggers, J.} \& \au{Bonn, D.}}
  \yr{2024{\natexlab{{\em b\/}}}}  \at{When does the elastic regime begin in
  viscoelastic pinch-off?}  \jt{arXiv preprint arXiv:2406.02303} .

\bibitem[Ga{\~n}{\'a}n-Calvo(2017)]{ganan2017revision}
{\sc \au{Ga{\~n}{\'a}n-Calvo, A.~M.}} \yr{2017}  \at{Revision of bubble
  bursting: Universal scaling laws of top jet drop size and speed}.  \jt{Phys.
  Rev. Lett.}  \bvol{119}~(20),  \pg{204502}.

\bibitem[Ghabache \& S{\'e}on(2016)]{ghabache2016size}
{\sc \au{Ghabache, {\'E}.} \& \au{S{\'e}on, T.}} \yr{2016}  \at{Size of the top
  jet drop produced by bubble bursting}.  \jt{Phys. Rev. Fluids}  \bvol{1}~(5),
   \pg{051901}.

\bibitem[Ghabache {\em et~al.\/}(2014)Ghabache, S{\'e}on \&
  Antkowiak]{ghabache2014liquid}
{\sc \au{Ghabache, {\'E}.}, \au{S{\'e}on, T.} \& \au{Antkowiak, A.}} \yr{2014}
  \at{Liquid jet eruption from hollow relaxation}.  \jt{J. Fluid Mech.}
  \bvol{761},  \pg{206--219}.

\bibitem[Giesekus(1982)]{giesekus1982simple}
{\sc \au{Giesekus, H.}} \yr{1982}  \at{A simple constitutive equation for
  polymer fluids based on the concept of deformation-dependent tensorial
  mobility}.  \jt{J. Non-Newton. Fluid Mech.}  \bvol{11}~(1-2),  \pg{69--109}.

\bibitem[Gonnermann \& Manga(2007)]{gonnermann2007fluid}
{\sc \au{Gonnermann, H.~M.} \& \au{Manga, M.}} \yr{2007}  \at{The fluid
  mechanics inside a volcano}.  \jt{Annu. Rev. Fluid Mech.}  \bvol{39},
  \pg{321--356}.

\bibitem[Gordillo \& Blanco-Rodr{\'\i}guez(2023)]{gordillo2023theory}
{\sc \au{Gordillo, J.~M.} \& \au{Blanco-Rodr{\'\i}guez, F.~J.}} \yr{2023}
  \at{Theory of the jets ejected after the inertial collapse of cavities with
  applications to bubble bursting jets}.  \jt{Phys. Rev. Fluids}  \bvol{8}~(7),
   \pg{073606}.

\bibitem[Gordillo {\em et~al.\/}(2020)Gordillo, Onuki \&
  Tagawa]{gordillo2020impulsive}
{\sc \au{Gordillo, J.~M.}, \au{Onuki, H.} \& \au{Tagawa, Y.}} \yr{2020}
  \at{Impulsive generation of jets by flow focusing}.  \jt{J. Fluid Mech.}
  \bvol{894}.

\bibitem[Gordillo \&
  Rodr{\'\i}guez-Rodr{\'\i}guez(2019)]{gordillo2019capillary}
{\sc \au{Gordillo, J.~M.} \& \au{Rodr{\'\i}guez-Rodr{\'\i}guez, J.}} \yr{2019}
  \at{Capillary waves control the ejection of bubble bursting jets}.  \jt{J.
  Fluid Mech.}  \bvol{867},  \pg{556--571}.

\bibitem[Goren \& Gottlieb(1982)]{goren1982surface}
{\sc \au{Goren, S.~L.} \& \au{Gottlieb, M.}} \yr{1982}
  \at{Surface-tension-driven breakup of viscoelastic liquid threads}.  \jt{J.
  Fluid Mech.}  \bvol{120},  \pg{245--266}.

\bibitem[Hinch(1993)]{hinch1993flow}
{\sc \au{Hinch, E.~J.}} \yr{1993}  \at{The flow of an oldroyd fluid around a
  sharp corner}.  \jt{J. Non-Newtonian Fluid Mech.}  \bvol{50}~(2-3),
  \pg{161--171}.

\bibitem[Hinch \& Harlen(2021)]{hinch2021oldroyd}
{\sc \au{Hinch, J.} \& \au{Harlen, O.}} \yr{2021}  \at{{Oldroyd B, and not A?}}
   \jt{J. Non-Newton. Fluid Mech.}  \bvol{298},  \pg{104668}.

\bibitem[Hinch {\em et~al.\/}(2024)Hinch, Boyko \& Stone]{hinch2024fast}
{\sc \au{Hinch, J.~E.}, \au{Boyko, E.} \& \au{Stone, H.~A.}} \yr{2024}
  \at{Fast flow of an oldroyd-b model fluid through a narrow slowly varying
  contraction}.  \jt{J. Fluid Mech.}  \bvol{988},  \pg{A11}.

\bibitem[Hosokawa {\em et~al.\/}(2023)Hosokawa, Kamamoto, Watanabe, Kusuno,
  Kobayashi \& Tagawa]{hosokawa2023phase}
{\sc \au{Hosokawa, A.}, \au{Kamamoto, K.}, \au{Watanabe, H.}, \au{Kusuno, H.},
  \au{Kobayashi, K.~U.} \& \au{Tagawa, Y.}} \yr{2023}  \at{A phase diagram of
  the pinch-off-type behavior of impulsively-induced viscoelastic liquid jets}.
   \jt{arXiv preprint arXiv:2309.01364} .

\bibitem[Ji {\em et~al.\/}(2023)Ji, Yang, Wang, Ewoldt \&
  Feng]{ji2023secondary}
{\sc \au{Ji, B.}, \au{Yang, Z.}, \au{Wang, Z.}, \au{Ewoldt, R.~H.} \& \au{Feng,
  J.}} \yr{2023}  \at{Secondary bubble entrainment via primary bubble bursting
  at a viscoelastic surface}.  \jt{Phys. Rev. Lett.}  \bvol{131}~(10),
  \pg{104002}.

\bibitem[Kant {\em et~al.\/}(2023)Kant, Pairetti, Saade, Popinet, Zaleski \&
  Lohse]{kant2023bag}
{\sc \au{Kant, P.}, \au{Pairetti, C.}, \au{Saade, Y.}, \au{Popinet, S.},
  \au{Zaleski, S.} \& \au{Lohse, D.}} \yr{2023}  \at{Bag-mediated film
  atomization in a cough machine}.  \jt{Phys. Rev. Fluids}  \bvol{8}~(7),
  \pg{074802}.

\bibitem[Kayal {\em et~al.\/}(2024)Kayal, Sanjay, Yewale, Kumar \&
  Dasgupta]{kayal2024focusing}
{\sc \au{Kayal, L.}, \au{Sanjay, V.}, \au{Yewale, N.}, \au{Kumar, A.} \&
  \au{Dasgupta, R.}} \yr{2024}  \at{Focusing of concentric free-surface waves}.
   \jt{arXiv preprint arXiv:2406.05416} .

\bibitem[Keller {\em et~al.\/}(1995)Keller, King \& Ting]{keller1995blob}
{\sc \au{Keller, J.~B.}, \au{King, A.} \& \au{Ting, L.}} \yr{1995}  \at{Blob
  formation}.  \jt{Phys. Fluids}  \bvol{7}~(1),  \pg{226--228}.

\bibitem[Kientzler {\em et~al.\/}(1954)Kientzler, Arons, Blanchard \&
  Woodcock]{kientzler1954photographic}
{\sc \au{Kientzler, C.~F.}, \au{Arons, A.~B.}, \au{Blanchard, D.~C.} \&
  \au{Woodcock, A.~H.}} \yr{1954}  \at{Photographic investigation of the
  projection of droplets by bubbles bursting at a water surface}.  \jt{Tellus}
  \bvol{6}~(1),  \pg{1--7}.

\bibitem[Knelman {\em et~al.\/}(1954)Knelman, Dombrowski \&
  Newitt]{knelman1954mechanism}
{\sc \au{Knelman, F.}, \au{Dombrowski, N.} \& \au{Newitt, D.~M.}} \yr{1954}
  \at{Mechanism of the bursting of bubbles}.  \jt{Nature}  \bvol{173}~(4397),
  \pg{261--261}.

\bibitem[Krishnan {\em et~al.\/}(2017)Krishnan, Hopfinger \&
  Puthenveettil]{krishnan2017scaling}
{\sc \au{Krishnan, S.}, \au{Hopfinger, E.~J.} \& \au{Puthenveettil, B.~A.}}
  \yr{2017}  \at{On the scaling of jetting from bubble collapse at a liquid
  surface}.  \jt{J. Fluid Mech.}  \bvol{822},  \pg{791}.

\bibitem[Le~Merrer {\em et~al.\/}(2012)Le~Merrer, Qu{\'e}r{\'e} \&
  Clanet]{le2012buckling}
{\sc \au{Le~Merrer, M.}, \au{Qu{\'e}r{\'e}, D.} \& \au{Clanet, C.}} \yr{2012}
  \at{Buckling of viscous filaments of a fluid under compression stresses}.
  \jt{Phys. Rev. Lett.}  \bvol{109}~(6),  \pg{064502}.

\bibitem[Lee \& Dalnoki-Veress(2024)]{lee2024buckling}
{\sc \au{Lee, C.~L.} \& \au{Dalnoki-Veress, K.}} \yr{2024}  \at{Buckling
  instability in a chain of sticky bubbles}.  \jt{Phys. Rev. Res.}
  \bvol{6}~(2),  \pg{L022062}.

\bibitem[Lee {\em et~al.\/}(2011)Lee, Weon, Park, Je, Fezzaa \&
  Lee]{lee2011size}
{\sc \au{Lee, J.~S.}, \au{Weon, B.~M.}, \au{Park, S.~J.}, \au{Je, J.~H.},
  \au{Fezzaa, K.} \& \au{Lee, W.~K.}} \yr{2011}  \at{Size limits the formation
  of liquid jets during bubble bursting}.  \jt{Nat. Commun.}  \bvol{2}~(1),
  \pg{367}.

\bibitem[Lhuissier \& Villermaux(2012)]{lhuissier2012bursting}
{\sc \au{Lhuissier, H.} \& \au{Villermaux, E.}} \yr{2012}  \at{Bursting bubble
  aerosols}.  \jt{J. Fluid Mech.}  \bvol{696},  \pg{5--44}.

\bibitem[Liger-Belair(2012)]{liger2012physics}
{\sc \au{Liger-Belair, G.}} \yr{2012}  \at{The physics behind the fizz in
  champagne and sparkling wines}.  \jt{Eur. Phys. J. Spec. Top.}
  \bvol{201}~(1),  \pg{1--88}.

\bibitem[Lin(1970)]{lin1970mechanisms}
{\sc \au{Lin, T.~J.}} \yr{1970}  \at{Mechanisms and control of gas bubble
  formation in cosmetics}.  \jt{J. Soc. Cosmet. Chem}  \bvol{22}~(6),
  \pg{323--337}.

\bibitem[Lohse(2003)]{lohse2003bubble}
{\sc \au{Lohse, D.}} \yr{2003}  \at{Bubble puzzles}.  \jt{Physics Today}
  \bvol{56}~(2),  \pg{36--41}.

\bibitem[Lohse(2018)]{Lohse2018}
{\sc \au{Lohse, D.}} \yr{2018}  \at{Bubble puzzles: From fundamentals to
  applications}.  \jt{Phys. Rev. Fluids.}  \bvol{3}~(11),  \pg{110504}.

\bibitem[Lohse(2022)]{lohse2022fundamental}
{\sc \au{Lohse, D.}} \yr{2022}  \at{Fundamental fluid dynamics challenges in
  inkjet printing}.  \jt{Annu. Rev. Fluid Mech.}  \bvol{54},  \pg{349--382}.

\bibitem[Lohse {\em et~al.\/}(2004)Lohse, Bergmann, Mikkelsen, Zeilstra, {van
  der Meer}, Versluis, {van der Weele}, {van der Hoef} \&
  Kuipers]{lohse2004impact}
{\sc \au{Lohse, D.}, \au{Bergmann, R.}, \au{Mikkelsen, R.}, \au{Zeilstra, C.},
  \au{{van der Meer}, D.}, \au{Versluis, M.}, \au{{van der Weele}, K.},
  \au{{van der Hoef}, M.} \& \au{Kuipers, H.}} \yr{2004}  \at{Impact on soft
  sand: void collapse and jet formation}.  \jt{Phys. Rev. Lett.}
  \bvol{93}~(19),  \pg{198003}.

\bibitem[Lohse \& Villermaux(2020)]{lohse-2020-pnas}
{\sc \au{Lohse, D.} \& \au{Villermaux, E.}} \yr{2020}  \at{Double threshold
  behavior for breakup of liquid sheets}.  \jt{Proc. Natl. Acad. Sci. U.S.A.}
  \bvol{117},  \pg{18912--18914}.

\bibitem[L{\'o}pez-Herrera {\em et~al.\/}(2019)L{\'o}pez-Herrera, Popinet \&
  Castrej{\'o}n-Pita]{lopez2019adaptive}
{\sc \au{L{\'o}pez-Herrera, J.-M.}, \au{Popinet, S.} \& \au{Castrej{\'o}n-Pita,
  A.-A.}} \yr{2019}  \at{An adaptive solver for viscoelastic incompressible
  two-phase problems applied to the study of the splashing of weakly
  viscoelastic droplets}.  \jt{J. Non-Newton. Fluid Mech.}  \bvol{264},
  \pg{144--158}.

\bibitem[{Lord Rayleigh}(1878)]{rayleigh1878instability}
{\sc \au{{Lord Rayleigh}}} \yr{1878}  \at{On the instability of jets}.
  \jt{Proc. Lond. Math. Soc.}  \bvol{1}~(1),  \pg{4--13}.

\bibitem[{Lord Rayleigh}(1896)]{rayleigh1896theory}
{\sc \au{{Lord Rayleigh}}} \yr{1896} {\em The Theory of Sound\/}.
  \publ{Dover}.

\bibitem[MacIntyre(1972)]{macintyre1972flow}
{\sc \au{MacIntyre, F.}} \yr{1972}  \at{Flow patterns in breaking bubbles}.
  \jt{J. Geophys. Res.}  \bvol{77}~(27),  \pg{5211--5228}.

\bibitem[Marchand {\em et~al.\/}(2011)Marchand, Weijs, Snoeijer \&
  Andreotti]{marchand2011surface}
{\sc \au{Marchand, A.}, \au{Weijs, J.~H.}, \au{Snoeijer, J.~H.} \&
  \au{Andreotti, B.}} \yr{2011}  \at{Why is surface tension a force parallel to
  the interface?}  \jt{Am. J. Phys.}  \bvol{79}~(10),  \pg{999--1008}.

\bibitem[Mason(1954)]{mason1954bursting}
{\sc \au{Mason, B.~J.}} \yr{1954}  \at{Bursting of air bubbles at the surface
  of sea water}.  \jt{Nature}  \bvol{174}~(4427),  \pg{470}.

\bibitem[Mathijssen {\em et~al.\/}(2023)Mathijssen, Lisicki, Prakash \&
  Mossige]{mathijssen2023culinary}
{\sc \au{Mathijssen, A. J. T.~M.}, \au{Lisicki, M.}, \au{Prakash, V.~N.} \&
  \au{Mossige, E. J.~L.}} \yr{2023}  \at{Culinary fluid mechanics and other
  currents in food science}.  \jt{Rev. Mod. Phys.}  \bvol{95}~(2),
  \pg{025004}.

\bibitem[Matoz-Fernandez {\em et~al.\/}(2020)Matoz-Fernandez, Davidson,
  Stanley-Wall \& Sknepnek]{matoz2020wrinkle}
{\sc \au{Matoz-Fernandez, D.~A.}, \au{Davidson, F.~A.}, \au{Stanley-Wall,
  N.~R.} \& \au{Sknepnek, R.}} \yr{2020}  \at{Wrinkle patterns in active
  viscoelastic thin sheets}.  \jt{Phys. Rev. Res.}  \bvol{2}~(1),  \pg{013165}.

\bibitem[McKinley \& Sridhar(2002)]{mckinley2002filament}
{\sc \au{McKinley, G.~H.} \& \au{Sridhar, T.}} \yr{2002}
  \at{Filament-stretching rheometry of complex fluids}.  \jt{Annu. Rev. Fluid
  Mech.}  \bvol{34}~(1),  \pg{375--415}.

\bibitem[Middleman(1965)]{middleman1965stability}
{\sc \au{Middleman, S.}} \yr{1965}  \at{Stability of a viscoelastic jet}.
  \jt{Chem. Eng. Sci.}  \bvol{20}~(12),  \pg{1037--1040}.

\bibitem[Moschopoulos {\em et~al.\/}(2021)Moschopoulos, Spyridakis, Varchanis,
  Dimakopoulos \& Tsamopoulos]{moschopoulos2021concept}
{\sc \au{Moschopoulos, P.}, \au{Spyridakis, A.}, \au{Varchanis, S.},
  \au{Dimakopoulos, Y.} \& \au{Tsamopoulos, J.}} \yr{2021}  \at{The concept of
  elasto-visco-plasticity and its application to a bubble rising in yield
  stress fluids}.  \jt{J. Non-Newton. Fluid Mech.}  \bvol{297},  \pg{104670}.

\bibitem[Munro(2019)]{munro2019coalescence}
{\sc \au{Munro, J.}} \yr{2019}  \at{Coalescence of bubbles and drops}. PhD
  thesis, {Univeristy of Cambridge}.

\bibitem[Oldroyd(1950)]{oldroyd1950formulation}
{\sc \au{Oldroyd, J.~G.}} \yr{1950}  \at{On the formulation of rheological
  equations of state}.  \jt{Proc. R. Soc. Lond.}  \bvol{200}~(1063),
  \pg{523--541}.

\bibitem[Oratis {\em et~al.\/}(2023)Oratis, Bertin \&
  Snoeijer]{oratis2023coalescence}
{\sc \au{Oratis, A.~T.}, \au{Bertin, V.} \& \au{Snoeijer, J.~H.}} \yr{2023}
  \at{Coalescence of bubbles in a viscoelastic liquid}.  \jt{Phys. Rev. Fluids}
   \bvol{8}~(8),  \pg{083603}.

\bibitem[Oratis {\em et~al.\/}(2020)Oratis, Bush, Stone \& Bird]{oratis2020new}
{\sc \au{Oratis, A.~T.}, \au{Bush, J. W.~M.}, \au{Stone, H.~A.} \& \au{Bird,
  J.~C.}} \yr{2020}  \at{A new wrinkle on liquid sheets: Turning the mechanism
  of viscous bubble collapse upside down}.  \jt{Science}  \bvol{369}~(6504),
  \pg{685--688}.

\bibitem[Oratis {\em et~al.\/}(2024)Oratis, Dijs, Lajoinie, Versluis \&
  Snoeijer]{oratis2024unifying}
{\sc \au{Oratis, A.~T.}, \au{Dijs, K.}, \au{Lajoinie, G.}, \au{Versluis, M.} \&
  \au{Snoeijer, J.~H.}} \yr{2024}  \at{A unifying rayleigh-plesset-type
  equation for bubbles in viscoelastic media}.  \jt{J. Acoust. Soc. Am.}
  \bvol{155}~(2),  \pg{1593--1605}.

\bibitem[Pandey {\em et~al.\/}(2021)Pandey, Kansal, Herrada, Eggers \&
  Snoeijer]{pandey2021elastic}
{\sc \au{Pandey, A.}, \au{Kansal, M.}, \au{Herrada, M.~A.}, \au{Eggers, J.} \&
  \au{Snoeijer, J.~H.}} \yr{2021}  \at{Elastic {Rayleigh--Plateau} instability:
  dynamical selection of nonlinear states}.  \jt{{Soft Matter}}
  \bvol{17}~(20),  \pg{5148--5161}.

\bibitem[Pico {\em et~al.\/}(2024)Pico, Kahouadji, Shin, Chergui, Juric \&
  Matar]{pico2024drop}
{\sc \au{Pico, P.~P.}, \au{Kahouadji, L.~L.}, \au{Shin, S.~S.}, \au{Chergui,
  J.~J.}, \au{Juric, D.~D.} \& \au{Matar, O.~K.}} \yr{2024}  \at{Drop
  encapsulation and bubble bursting in surfactant-laden flows in capillary
  channels}.  \jt{Phys. Rev. Fluids}  \bvol{9}~(3),  \pg{034001}.

\bibitem[Pierre {\em et~al.\/}(2022)Pierre, Poujol \&
  S{\'e}on]{pierre2022influence}
{\sc \au{Pierre, J.}, \au{Poujol, M.} \& \au{S{\'e}on, T.}} \yr{2022}
  \at{Influence of surfactant concentration on drop production by bubble
  bursting}.  \jt{Phys. Rev. Fluids}  \bvol{7}~(7),  \pg{073602}.

\bibitem[Plateau(1873)]{plateau1873statique}
{\sc \au{Plateau, J. A.~F.}} \yr{1873} {\em Statique exp{\'e}rimentale et
  th{\'e}orique des liquides soumis aux seules forces mol{\'e}culaires: Tome
  premier\/}, ,  \vol{vol.~2}.  \publ{Gauthier-Villars}.

\bibitem[Popinet(2009)]{popinet2009accurate}
{\sc \au{Popinet, S.}} \yr{2009}  \at{An accurate adaptive solver for
  surface-tension-driven interfacial flows}.  \jt{J. Comput. Phys.}
  \bvol{228}~(16),  \pg{5838--5866}.

\bibitem[Popinet(2015)]{popinet2015quadtree}
{\sc \au{Popinet, S.}} \yr{2015}  \at{A quadtree-adaptive multigrid solver for
  the serre--green--naghdi equations}.  \jt{J. Comput. Phys.}  \bvol{302},
  \pg{336--358}.

\bibitem[Popinet(2018)]{popinet2018numerical}
{\sc \au{Popinet, S.}} \yr{2018}  \at{Numerical models of surface tension}.
  \jt{Annu. Rev. Fluid Mech.}  \bvol{50},  \pg{49--75}.

\bibitem[Popinet \& {collaborators}(2013--2024)]{basilliskpopinet}
{\sc \au{Popinet, S.} \& \au{{collaborators}}} \yr{2013--2024} Basilisk {C}.
  \url{http://basilisk.fr} (Last accessed: June, 2024).

\bibitem[Princen(1963)]{princen1963shape}
{\sc \au{Princen, H.~M.}} \yr{1963}  \at{Shape of a fluid drop at a
  liquid-liquid interface}.  \jt{J. Colloid Sci.}  \bvol{18}~(2),
  \pg{178--195}.

\bibitem[Putz \& Burghelea(2009)]{putz2009solid}
{\sc \au{Putz, A. M.~V.} \& \au{Burghelea, T.~I.}} \yr{2009}  \at{The
  solid--fluid transition in a yield stress shear thinning physical gel}.
  \jt{Rheol. Acta}  \bvol{48},  \pg{673--689}.

\bibitem[Remmelgas {\em et~al.\/}(1999)Remmelgas, Singh \&
  Leal]{remmelgas1999computational}
{\sc \au{Remmelgas, J.}, \au{Singh, P.} \& \au{Leal, L.~G.}} \yr{1999}
  \at{Computational studies of nonlinear elastic dumbbell models of boger
  fluids in a cross-slot flow}.  \jt{J. Non-Newton. Fluid Mech.}
  \bvol{88}~(1-2),  \pg{31--61}.

\bibitem[Renardy \& Thomases(2021)]{renardy2021mathematician}
{\sc \au{Renardy, M.} \& \au{Thomases, B.}} \yr{2021}  \at{A mathematician's
  perspective on the oldroyd b model: progress and future challenges}.  \jt{J.
  Non-Newton. Fluid Mech.}  \bvol{293},  \pg{104573}.

\bibitem[Rodr{\'\i}guez-D{\'\i}az {\em et~al.\/}(2023)Rodr{\'\i}guez-D{\'\i}az,
  Rubio, Montanero, Ga{\~n}{\'a}n-Calvo \& Cabezas]{rodriguez2023bubble}
{\sc \au{Rodr{\'\i}guez-D{\'\i}az, P.}, \au{Rubio, A.}, \au{Montanero, J.~M.},
  \au{Ga{\~n}{\'a}n-Calvo, A.~M.} \& \au{Cabezas, M.~G.}} \yr{2023}  \at{Bubble
  bursting in a weakly viscoelastic liquid}.  \jt{Phys. Fluids}
  \bvol{35}~(10).

\bibitem[Sanjay(2022)]{VatsalThesis}
{\sc \au{Sanjay, V.}} \yr{2022}  \at{{Viscous Free-Surface Flows}}. PhD thesis,
  University of Twente.

\bibitem[Sanjay(2024)]{vatsalElastoFlow2024}
{\sc \au{Sanjay, V.}} \yr{2024} Code repository: {Basilisk C ElastoFlow} -
  {C}omplete {2D/3D} viscoelastic framework.
  \href{https://doi.org/10.5281/zenodo.14210635}{10.5281/zenodo.14210635}.

\bibitem[Sanjay \& Dixit(2024)]{Sanjay2024code}
{\sc \au{Sanjay, V.} \& \au{Dixit, A.~K.}} \yr{2024} Code repository:
  {Viscoelastic Worthington Jets \& Droplets Produced by Bursting Bubbles}.
  \href{https://doi.org/10.5281/zenodo.14349207}{10.5281/zenodo.14349207}.

\bibitem[Sanjay {\em et~al.\/}(2021)Sanjay, Lohse \&
  Jalaal]{sanjay2021bursting}
{\sc \au{Sanjay, V.}, \au{Lohse, D.} \& \au{Jalaal, M.}} \yr{2021}
  \at{Bursting bubble in a viscoplastic medium}.  \jt{J. Fluid Mech.}
  \bvol{922},  \pg{A2}.

\bibitem[Sanjay {\em et~al.\/}(2022)Sanjay, Sen, Kant \&
  Lohse]{sanjay-2022-JFM}
{\sc \au{Sanjay, V.}, \au{Sen, U.}, \au{Kant, P.} \& \au{Lohse, D.}} \yr{2022}
  \at{Taylor-{C}ulick retractions and the influence of the surroundings}.
  \jt{J. Fluid Mech.}  \bvol{948},  \pg{A14}.

\bibitem[Saramito(2007)]{saramito2007}
{\sc \au{Saramito, P.}} \yr{2007}  \at{A new constitutive equation for
  elastoviscoplastic fluid flows}.  \jt{J. Non-Newtonian Fluid Mech.}
  \bvol{145}~(1),  \pg{1--14}.

\bibitem[Schmalholz \& Podladchikov(1999)]{schmalholz1999buckling}
{\sc \au{Schmalholz, S.~M.} \& \au{Podladchikov, Y.}} \yr{1999}  \at{Buckling
  versus folding: importance of viscoelasticity}.  \jt{Geophys. Res. Lett.}
  \bvol{26}~(17),  \pg{2641--2644}.

\bibitem[Sen {\em et~al.\/}(2021)Sen, Datt, Segers, Wijshoff, Snoeijer,
  Versluis \& Lohse]{sen2021retraction}
{\sc \au{Sen, U.}, \au{Datt, C.}, \au{Segers, T.}, \au{Wijshoff, H.},
  \au{Snoeijer, J.~H.}, \au{Versluis, M.} \& \au{Lohse, D.}} \yr{2021}  \at{The
  retraction of jetted slender viscoelastic liquid filaments}.  \jt{J. Fluid
  Mech.}  \bvol{929},  \pg{A25}.

\bibitem[Sen {\em et~al.\/}(2024)Sen, Zinelis, Sanjay, Matar, Lohse \&
  Jalaal]{sen2024elastocapillary}
{\sc \au{Sen, U.}, \au{Zinelis, K.}, \au{Sanjay, V.}, \au{Matar, O.~K.},
  \au{Lohse, Detlef} \& \au{Jalaal, Maziyar}} \yr{2024}  \at{Elastocapillary
  worthington jets}.  \jt{arXiv preprint arXiv:2207.07928} .

\bibitem[Shi {\em et~al.\/}(1994)Shi, Brenner \& Nagel]{shi1994cascade}
{\sc \au{Shi, X.~D.}, \au{Brenner, M.~P.} \& \au{Nagel, S.~R.}} \yr{1994}
  \at{A cascade of structure in a drop falling from a faucet}.  \jt{Science}
  \bvol{265}~(5169),  \pg{219--222}.

\bibitem[Singh \& Das(2019)]{singh2019numerical}
{\sc \au{Singh, D.} \& \au{Das, A.~K.}} \yr{2019}  \at{Numerical investigation
  of the collapse of a static bubble at the free surface in the presence of
  neighbors}.  \jt{Phys. Rev. Fluids}  \bvol{4}~(2),  \pg{023602}.

\bibitem[Singh \& Das(2021)]{singh2021dynamics}
{\sc \au{Singh, D.} \& \au{Das, A.~K.}} \yr{2021}  \at{Dynamics of inner gas
  during the bursting of a bubble at the free surface}.  \jt{Phys. Fluids}
  \bvol{33}~(5).

\bibitem[Snoeijer {\em et~al.\/}(2020)Snoeijer, Pandey, Herrada \&
  Eggers]{snoeijer2020relationship}
{\sc \au{Snoeijer, J.~H.}, \au{Pandey, A.}, \au{Herrada, M.~A.} \& \au{Eggers,
  J.}} \yr{2020}  \at{The relationship between viscoelasticity and elasticity}.
   \jt{Proc. R. Soc. A}  \bvol{476}~(2243),  \pg{20200419}.

\bibitem[Stokes(1845)]{stokes1845}
{\sc \au{Stokes, G.~G.}} \yr{1845}  \at{On the theories of the internal
  friction of fluids in motion, and of the equilibrium and motion of elastic
  solids}.  \jt{Trans. Cambridge Philos. Soc.}  \bvol{8},  \pg{287}.

\bibitem[Stone \& Leal(1989)]{stone1989relaxation}
{\sc \au{Stone, H.~A.} \& \au{Leal, L.~G.}} \yr{1989}  \at{Relaxation and
  breakup of an initially extended drop in an otherwise quiescent fluid}.
  \jt{J. Fluid Mech.}  \bvol{198},  \pg{399--427}.

\bibitem[Stone {\em et~al.\/}(2023)Stone, Shelley \& Boyko]{stone2023note}
{\sc \au{Stone, H.~A.}, \au{Shelley, M.~J.} \& \au{Boyko, E.}} \yr{2023}  \at{A
  note about convected time derivatives for flows of complex fluids}.
  \jt{{Soft Matter}}  \bvol{19}~(28),  \pg{5353--5359}.

\bibitem[{Stuhlman Jr}(1932)]{stuhlman1932mechanics}
{\sc \au{{Stuhlman Jr}, O.}} \yr{1932}  \at{The mechanics of effervescence}.
  \jt{Physics}  \bvol{2}~(6),  \pg{457--466}.

\bibitem[Tanner(2000)]{tanner2000engineering}
{\sc \au{Tanner, R.~I.}} \yr{2000} {\em Engineering rheology\/}, ,
  \vol{vol.~52}.  \publ{OUP Oxford}.

\bibitem[Taylor(1959)]{taylor-1959-procrsoclonda}
{\sc \au{Taylor, G.~I.}} \yr{1959}  \at{The dynamics of thin sheets of fluid.
  {III}. {D}isintegration of fluid sheets}.  \jt{Proc. R. Soc. Lond.}
  \bvol{253},  \pg{313--321}.

\bibitem[Taylor(1969)]{taylor1969instability}
{\sc \au{Taylor, G.~I.}} \yr{1969} Instability of jets, threads, and sheets of
  viscous fluid.  \bt{In {\em Proceedings of the Twelfth International Congress
  of Applied Mechanics, Stanford\/}},  \pg{pp. 382--388}. Springer.

\bibitem[Timoshenko \& Gere(2012)]{timoshenko2012theory}
{\sc \au{Timoshenko, S.~P.} \& \au{Gere, J.~M.}} \yr{2012} {\em Theory of
  elastic stability\/}.  \publ{Courier Corporation}.

\bibitem[Toba(1959)]{toba1959drop}
{\sc \au{Toba, Y.}} \yr{1959}  \at{Drop production by bursting of air bubbles
  on the sea surface (ii) theoretical study on the shape of floating bubbles}.
  \jt{J. Oceanogr. Soc. Jpn.}  \bvol{15}~(3),  \pg{121--130}.

\bibitem[Trouton(1906)]{trouton1906coefficient}
{\sc \au{Trouton, F.~T.}} \yr{1906}  \at{On the coefficient of viscous traction
  and its relation to that of viscosity}.  \jt{Proc. R. Soc. Lond.}
  \bvol{77}~(519),  \pg{426--440}.

\bibitem[Tryggvason {\em et~al.\/}(2011)Tryggvason, Scardovelli \&
  Zaleski]{tryggvason2011direct}
{\sc \au{Tryggvason, G.}, \au{Scardovelli, R.} \& \au{Zaleski, S.}} \yr{2011}
  {\em {Direct} {Numerical} {Simulations} of {Gas}--{Liquid} {Multiphase}
  {Flows}\/}.  \publ{Cambridge University Press}.

\bibitem[Turkoz {\em et~al.\/}(2018)Turkoz, Lopez-Herrera, Eggers, Arnold \&
  Deike]{turkoz2018axisymmetric}
{\sc \au{Turkoz, E.}, \au{Lopez-Herrera, J.~M.}, \au{Eggers, J.}, \au{Arnold,
  C.~B.} \& \au{Deike, L.}} \yr{2018}  \at{Axisymmetric simulation of
  viscoelastic filament thinning with the oldroyd-b model}.  \jt{J. Fluid
  Mech.}  \bvol{851},  \pg{R2}.

\bibitem[Turkoz {\em et~al.\/}(2021)Turkoz, Stone, Arnold \&
  Deike]{turkoz2021simulation}
{\sc \au{Turkoz, E.}, \au{Stone, H.~A.}, \au{Arnold, C.~B.} \& \au{Deike, L.}}
  \yr{2021}  \at{Simulation of impulsively induced viscoelastic jets using the
  oldroyd-b model}.  \jt{J. Fluid Mech.}  \bvol{911},  \pg{A14}.

\bibitem[Varchanis {\em et~al.\/}(2019)Varchanis, Makrigiorgos, Moschopoulos,
  Dimakopoulos \& Tsamopoulos]{varchanis2019modeling}
{\sc \au{Varchanis, S.}, \au{Makrigiorgos, G.}, \au{Moschopoulos, P.},
  \au{Dimakopoulos, Y.} \& \au{Tsamopoulos, J.}} \yr{2019}  \at{Modeling the
  rheology of thixotropic elasto-visco-plastic materials}.  \jt{J. Rheol.}
  \bvol{63}~(4),  \pg{609--639}.

\bibitem[Varchanis \& Tsamopoulos(2022)]{varchanis2022numerical}
{\sc \au{Varchanis, S.} \& \au{Tsamopoulos, J.}} \yr{2022}  \at{Numerical
  simulations of interfacial and elastic instabilities}.  \jt{{Science Talks}}
  \bvol{3},  \pg{100053}.

\bibitem[Villermaux {\em et~al.\/}(2022)Villermaux, Wang \&
  Deike]{villermaux2022bubbles}
{\sc \au{Villermaux, E.}, \au{Wang, X.} \& \au{Deike, L.}} \yr{2022}
  \at{Bubbles spray aerosols: certitudes and mysteries}.  \jt{{Proc. Natl.
  Acad. Sci. Nexus}}  \bvol{1}~(5),  \pg{pgac261}.

\bibitem[Walls {\em et~al.\/}(2015)Walls, Henaux \& Bird]{walls2015jet}
{\sc \au{Walls, P. L.~L.}, \au{Henaux, L.} \& \au{Bird, J.~C.}} \yr{2015}
  \at{Jet drops from bursting bubbles: How gravity and viscosity couple to
  inhibit droplet production}.  \jt{Phys. Rev. E}  \bvol{92}~(2),  \pg{021002}.

\bibitem[Walls {\em et~al.\/}(2017)Walls, McRae, Natarajan, Johnson, Antoniou
  \& Bird]{walls2017quantifying}
{\sc \au{Walls, P. L.~L.}, \au{McRae, O.}, \au{Natarajan, V.}, \au{Johnson,
  C.}, \au{Antoniou, C.} \& \au{Bird, J.~C.}} \yr{2017}  \at{Quantifying the
  potential for bursting bubbles to damage suspended cells}.  \jt{Sci. Rep.}
  \bvol{7}~(1),  \pg{15102}.

\bibitem[Woodcock {\em et~al.\/}(1953)Woodcock, Kientzler, Arons \&
  Blanchard]{woodcock1953giant}
{\sc \au{Woodcock, A.~H.}, \au{Kientzler, C.~F.}, \au{Arons, A.~B.} \&
  \au{Blanchard, D.~C.}} \yr{1953}  \at{Giant condensation nuclei from bursting
  bubbles}.  \jt{Nature}  \bvol{172}~(4390),  \pg{1144--1145}.

\bibitem[Worthington(1877)]{worthington1877xxviii}
{\sc \au{Worthington, A.~M.}} \yr{1877}  \at{{On} the forms assumed by drops of
  liquids falling vertically on a horizontal plate}.  \jt{Proc. R. Soc. Lond.}
  \bvol{25}~(171-178),  \pg{261--272}.

\bibitem[Worthington(1908)]{worthington1908study}
{\sc \au{Worthington, A.~M.}} \yr{1908} {\em A study of splashes\/}.
  \publ{London: Longman, Green and Co}.

\bibitem[Yamani \& McKinley(2023)]{yamani2023master}
{\sc \au{Yamani, S.} \& \au{McKinley, G.~H.}} \yr{2023}  \at{Master curves for
  {FENE-P} fluids in steady shear flow}.  \jt{J. Non-Newton. Fluid Mech.}
  \bvol{313},  \pg{104944}.

\bibitem[Yang {\em et~al.\/}(2023)Yang, Ji, Ault \& Feng]{yang2023enhanced}
{\sc \au{Yang, Z.}, \au{Ji, B.}, \au{Ault, J.~T.} \& \au{Feng, J.}} \yr{2023}
  \at{Enhanced singular jet formation in oil-coated bubble bursting}.  \jt{Nat.
  Phys.}  \bvol{19}~(6),  \pg{884--890}.

\bibitem[Yang {\em et~al.\/}(2020)Yang, Tian \& Thoroddsen]{yang2020multitude}
{\sc \au{Yang, Z.~Q.}, \au{Tian, Y.~S.} \& \au{Thoroddsen, S.~T.}} \yr{2020}
  \at{Multitude of dimple shapes can produce singular jets during the collapse
  of immiscible drop-impact craters}.  \jt{J. Fluid Mech.}  \bvol{904},
  \pg{A19}.

\bibitem[Yarin(1993)]{yarin1993free}
{\sc \au{Yarin, A.~L.}} \yr{1993} {\em Free liquid jets and films:
  hydrodynamics and rheology\/}.  \publ{Longman Scientific and Technical}.

\bibitem[Zeff {\em et~al.\/}(2000)Zeff, Kleber, Fineberg \&
  Lathrop]{zeff2000singularity}
{\sc \au{Zeff, B.~W.}, \au{Kleber, B.}, \au{Fineberg, J.} \& \au{Lathrop,
  D.~P.}} \yr{2000}  \at{Singularity dynamics in curvature collapse and jet
  eruption on a fluid surface}.  \jt{Nature}  \bvol{403}~(6768),
  \pg{401--404}.

\bibitem[Zinelis {\em et~al.\/}(2024)Zinelis, Abadie, McKinley \&
  Matar]{zinelis2023transition}
{\sc \au{Zinelis, K.}, \au{Abadie, T.}, \au{McKinley, G.~H.} \& \au{Matar,
  O.~K.}} \yr{2024}  \at{Transition to elasto-capillary thinning dynamics in
  viscoelastic jets}.  \jt{J. Fluid Mech.}  \bvol{998},  \pg{A4}.

\end{thebibliography}

\end{document}